\documentclass[english,11pt,aps,prd,a4paper,preprintnumbers,floatfix,nofootinbib,showpacs,superscriptaddress, notitlepage]{revtex4-1} 
\usepackage[mode=buildnew]{standalone}
\usepackage[usenames,dvipsnames]{color}  
\usepackage{graphicx}
\usepackage{setspace}
\usepackage{caption}
\usepackage{subcaption}
\usepackage{amsmath}
\usepackage{amssymb}
\usepackage{soul}
\usepackage[colorlinks=true,citecolor=darkred,urlcolor=darkred, pdfborder={0 0 0}]{hyperref}
\usepackage[normalem]{ulem}
\usepackage{xcolor}
\usepackage{placeins}
\usepackage{mathrsfs}
\usepackage{verbatim} 
\usepackage{cancel} 
\usepackage{float} 

\usepackage{scalerel}
\usepackage{tikz-feynman}
\tikzfeynmanset{compat=1.1.0}
\usepackage{feynmp}
\usepackage{tikzsymbols}
\usepackage{multirow}

\makeatletter
\def\p@subsection{}
\makeatother

\definecolor{darkred}{rgb}{0.6,0,0}

\definecolor{linkcolor}{rgb}{0,0,0.5}

\def\gsim{\raise0.3ex\hbox{$\;>$\kern-0.75em\raise-1.1ex\hbox{$\sim\;$}}}
\def\lsim{\raise0.3ex\hbox{$\;<$\kern-0.75em\raise-1.1ex\hbox{$\sim\;$}}}

\def\beqn#1{\begin{equation}\label{#1}}
\def\eeqn{\end{equation}}

\def\beqa#1{\begin{eqnarray}\label{#1}}
\def\eeqa{\end{eqnarray}}

\newcommand {\ignore}[1]{}

\def\Z4{$Z_4$}
\def\O5{$\mathcal{O}_5$ }

\def\321{$\mathrm{SU(3) \otimes SU(2) \otimes U(1)}$ }

\usepackage{orcidlink}

\begin{document}

\title{\color{Blue} Radiative Lifting of $\mathbb{Z}_3$ Domain-Wall Degeneracy in a Type-III Seesaw Model: Implications for Leptogenesis and Gravitational Waves}

 \author{Priya\orcidlink{0000-0003-1183-4441}
 }\email{kashyappriya963@gmail.com}
 \affiliation{Department of Physics and Astronomical Science, Central University of Himachal Pradesh
Dharamshala, India 176215}
\author{Labh Singh\orcidlink{0000-0001-7704-726X}
}\email{sainilabh5@gmail.com}
 \affiliation{Department of Physics and Astronomical Science, Central University of Himachal Pradesh
Dharamshala, India 176215}
\author{B. C. Chauhan\orcidlink{0000-0002-9355-9772}
}\email{bcawake@hpcu.ac.in}
\affiliation{Department of Physics and Astronomical Science, Central University of Himachal Pradesh
Dharamshala, India 176215}
\author{Surender Verma\orcidlink{0000-0002-5671-5369}
}\email{s\_7verma@hpcu.ac.in}
 \affiliation{Department of Physics and Astronomical Science, Central University of Himachal Pradesh
Dharamshala, India 176215}
\begin{abstract}
  \vspace{1cm} 
\noindent In this work, we study a $\mathbb{Z}_3$-symmetric extension of the Standard Model with three hyperchargeless $SU(2)_L$ fermion triplets responsible for neutrino mass generation $\textit{via}$ the Type-III seesaw mechanism together with a complex scalar singlet $\chi$ whose vacuum expectation value spontaneously breaks the $\mathbb{Z}_3$ symmetry. Radiative corrections induced by the Yukawa interactions between the $SU(2)_L$ fermion triplets and the complex scalar singlet $\chi$ generate a Coleman-Weinberg vacuum bias that lifts the degeneracy among the $\mathbb{Z}_3$ vacua, leading to the annihilation of unstable domain-walls. Consequently, the degeneracy among the $\mathbb{Z}_3$ vacua is lifted radiatively through the Coleman-Weinberg effective potential, generating a dynamical bias term that triggers the annihilation of unstable domain walls. We perform a numerical analysis consistent with current neutrino oscillation data and identify viable regions of parameter space accommodating the observed neutrino masses and leptonic mixing parameters. The observed baryon asymmetry of the Universe is generated through thermal leptogenesis $\textit{via}$ the out-of-equilibrium decay of the lightest fermion triplet for masses around $\mathcal{O}(10^{9})\,\mathrm{GeV}$, consistent with the Type-III seesaw framework. 
Depending on the choice of model parameters, the predicted gravitational-wave spectrum can fall within the sensitivity reach of future space-based and ground-based gravitational-wave detectors. Our framework therefore establishes a correlation between neutrino mass generation, leptogenesis, radiative domain-wall instability, and gravitational-wave phenomenology.
\end{abstract}
\maketitle


\section{Introduction}\label{section1} 
\noindent The Standard Model (SM) of particle physics successfully describes the known elementary particles and their interactions, where neutrinos are treated as massless left-handed Weyl fermions. However, results from neutrino oscillation experiments such as Super-Kamiokande~\cite{Super-Kamiokande:1998kpq}, the Sudbury Neutrino Observatory (SNO)~\cite{SNO:2002tuh} and KamLAND~\cite{KamLAND:2002uet} established that neutrinos possess tiny, however non-zero masses, providing evidence for physics beyond the SM. 
However, the origin of neutrino masses remains unresolved. In this context, the seesaw mechanism provides a natural explanation for the smallness of neutrino masses through the introduction of heavy new states that suppresses the effective light neutrino masses~\cite{Minkowski:1977sc, Mohapatra:1979ia, Foot:1988aq}. An important feature of this mechanism is that it can, also, account for the observed baryon asymmetry of the Universe. This is achieved through the out-of-equilibrium decay of the lightest fermion triplet which generates a lepton asymmetry that is subsequently converted into baryon asymmetry $\textit{via}$ sphaleron processes~\cite{Kuzmin:1985mm}.

\noindent In particle physics models, discrete symmetries are often introduced to realize specific phenomenological features~\cite{Tapender:2023kdk,Chauhan:2023faf,Barman:2025bru,Priya:2025khf,Palavric:2024gvu}. When such symmetries are spontaneously broken in the early Universe, they can lead to the formation of stable domain-walls. The cosmological evolution and eventual decay of these topological defects can generate stochastic backgrounds of gravitational waves (GWs) making them particularly relevant in the era of gravitational-wave astronomy. Since the first direct detection of gravitational waves by the LIGO and Virgo Collaborations~\cite{LIGOScientific:2016aoc,LIGOScientific:2017vwq}, gravitational-wave observations have emerged as a powerful probe of physics beyond the SM. In particular, gravitational waves originating from the early Universe can provide valuable information about high-scale symmetry-breaking phenomena and the associated vacuum structure of new physics models.

\noindent However, stable domain-walls, also, lead to a serious cosmological problem. Their energy density decreases more slowly than that of radiation or matter causing them to eventually dominate the energy density of the Universe and potentially overclose it. A well-known way to evade this problem is to introduce a small explicit breaking of the discrete symmetry in addition to its spontaneous breaking. Such explicit symmetry-breaking effects lift the degeneracy among the vacua, thereby destabilizing the domain-walls and allowing them to decay sufficiently early. In many conventional scenarios, this bias term is introduced phenomenologically through small explicit symmetry-breaking operators in the scalar potential, for example,
\begin{equation}
V(\chi) \supset -\mu^2 |\chi|^2 + \frac{\lambda}{4} |\chi|^4 + \left( \kappa \, \chi^3 + \text{h.c.} \right),
\end{equation}
where the cubic interaction explicitly breaks the discrete symmetry and induces the pressure difference required for domain-wall annihilation. However, for the generated gravitational-wave signal to be phenomenologically relevant, the explicit breaking must remain sufficiently small so that the domain-walls survive long enough to radiate gravitational waves while still decaying early enough to avoid cosmological problems. The production of gravitational waves from topological defects has been extensively investigated in the literature~\cite{Vilenkin:1984ib,Hiramatsu:2010yz,Hiramatsu:2013qaa,Kitajima:2015nla,Saikawa:2017hiv,Krajewski:2017czs,Chen:2020wvu,Borah:2026kfo,Ma:2025bjf,Paul:2024iie,Dey:2025pcs,Biswas:2025rzs,Roshan:2026yon,Gouttenoire:2025ofv,Roshan:2024qnv,Bhattacharya:2023kws,Wu:2022stu,Barreto:2026igt,Li:2025gld,Fu:2024jhu,Wu:2022tpe,Wei:2022poh,Jana:2025vyb,Blasi:2024vew,Bhandari:2026ujy,Borah:2025bfa,Dror:2019syi,Blasi:2020wpy,Fornal:2020esl,Samanta:2020cdk,Barman:2022yos,Huang:2022vkf, Dasgupta:2022isg,Okada:2018xdh, Hasegawa:2019amx,Borah:2022cdx,Borah:2023saq,Borah:2022vsu,Barman:2023fad,Chen:2026fod,King:2024lki,Barreto:2026oyf,Winckler:2025hbc,Zeng:2025zjp,Kibble:1976sj,Zeldovich:1974uw}.

\noindent Although unstable domain-wall dynamics have been extensively studied in the context of a simpler $\mathbb{Z}_2$ symmetry~\cite{Vilenkin:1984ib, Hiramatsu:2010yz,Hiramatsu:2013qaa, Kitajima:2015nla, Saikawa:2017hiv, Krajewski:2017czs, Chen:2020wvu, Borah:2026kfo, Ma:2025bjf, Roshan:2026yon,Gouttenoire:2025ofv, Wu:2022stu,Barreto:2026igt, Li:2025gld, Fu:2024jhu, Wu:2022tpe, Wei:2022poh,Bhandari:2026ujy,Borah:2025bfa, Barman:2022yos}, in the present work, we focus on the richer phenomenology associated with a $\mathbb{Z}_3$ symmetry. In particular, the spontaneous breaking of the $\mathbb{Z}_3$ symmetry gives rise to three degenerate vacua connected through non-trivial complex phases leading to a more intricate domain-wall configuration and vacuum structure compared to the $\mathbb{Z}_2$ case. Consequently, the instability of the domain-walls is directly tied to the same sector responsible for neutrino mass generation and leptogenesis. This allows us to explore the corresponding implications for neutrino mass generation, thermal leptogenesis, and gravitational-wave phenomenology within the Type-III seesaw framework. In most studies, the origin and smallness of this bias term are simply assumed rather than dynamically explained. In the present framework, the required $\mathbb{Z}_3$-breaking bias term arises naturally through one-loop quantum corrections involving the heavy fermion triplets of the Type-III seesaw sector. In this work, we consider the well-known Type-III seesaw framework extended by a complex scalar singlet $\chi$.
The neutrino masses are generated at tree level while the Lagrangian remains exactly $\mathbb{Z}_3$ . We further introduce explicit $\mathbb{Z}_3$-breaking Yukawa interactions between the scalar singlet $\chi$ and the fermion triplets $\Sigma_i$. These interactions generate the required bias term radiatively through the one-loop Coleman–Weinberg effective potential, naturally lifting the vacuum degeneracy without introducing any ad hoc symmetry-breaking term. Consequently, the instability of the domain-walls originates dynamically from the same fermionic sector responsible for neutrino mass generation. The subsequent annihilation of these unstable domain-walls gives rise to a stochastic gravitational-wave background. We analyze the resulting gravitational-wave spectrum across the relevant parameter space and show that, for suitable choices of the $\mathbb{Z}_3$-breaking scale, seesaw scale, and $\chi$–$\Sigma_i$ couplings, the predicted signal can lie within the sensitivity reach of future gravitational-wave detectors such as TianQin~\cite{TianQin:2015yph,TianQin:2020hid}, Taiji~\cite{Hu:2017mde, Ruan:2018tsw}, LISA~\cite{LISA:2017pwj}, BBO~\cite{Crowder:2005nr},ET~\cite{Punturo:2010zz,Hild:2010id,Sathyaprakash:2012jk}, LIGO~\cite{LIGOScientific:2014qfs,LIGOScientific:2016aoc,LIGOScientific:2016wof,LIGOScientific:2017vwq}, and CE~\cite{Reitze:2019iox}. Thus, the framework establishes a unified connection between neutrino mass generation, leptogenesis, domain-wall dynamics, and gravitational-wave phenomenology.

\noindent The paper is organized as follows: Section~\ref{section2} presents the field content of the model, and the relevant mass terms. Section~\ref{section3} describes the generation of neutrino masses through the Type-III seesaw mechanism. Section~\ref{section4} discusses the mechanism of leptogenesis and the resulting lepton asymmetry. Section~\ref{section5} analyzes the formation and evolution of domain-walls arising from the spontaneous breaking of the $\mathbb{Z}_3$ symmetry, including the effects of explicit symmetry breaking. Finally, the conclusions are presented in Section~\ref{section6}.
\section{Model and Formalism}\label{section2}
\begin{table}[t]
\centering
\begin{tabular}{ccccccc}
\hline
 & $L_e$, $L_\mu$, $L_\tau$ & $e_R$, $\mu_R$, $\tau_R$ & $\Sigma_1$, $\Sigma_2$, $\Sigma_3$ & $H$ & $\chi$  \\
\hline
$SU(2)$ & $2$ & $1$ & $3$ & $2$ & $1$ \\
$\mathbb{Z}_3$   & $\omega$ & $1$ & $1$ &$\omega$ &  $\omega^2$  \\
\hline
\end{tabular}
\caption{Field content and charge assignments of Type-III seesaw mechanism under $\mathbb{Z}_3$ group.}
\label{tab:charges}
\end{table}
\noindent In our framework, we consider a Type-III seesaw scenario containing three heavy fermionic triplets, denoted by $\Sigma_1$, $\Sigma_2$, and $\Sigma_3$. To restrict the Yukawa and mass structures, we impose a discrete $\mathbb{Z}_3$ symmetry under which the triplet fermions transform as $1$. In addition to the Standard Model (SM) particle content, we introduce a complex gauge-singlet scalar field, $\chi$, carrying a $\mathbb{Z}_3$ charge $\omega^2$, where $\omega=e^{2\pi i/3}$ satisfies $\omega^3=1$. The imposed symmetry and the extended scalar sector determine the allowed interaction terms and consequently the structure of the neutrino mass matrix. The particle content and their corresponding charge assignments are summarized in Table~\ref{tab:charges}. The corresponding invariant Yukawa Lagrangian is given as

\begin{equation}
\begin{split}
-\mathcal{L} \; = \;
& y_{ee}\,\bar{L}_e H e_R
+ y_{e\mu}\,\bar{L}_e H \mu_R
+ y_{e\tau}\,\bar{L}_e H \tau_R \\[2mm]
&+ y_{\mu e}\,\bar{L}_\mu H e_R
+ y_{\mu\mu}\,\bar{L}_\mu H \mu_R
+ y_{\mu\tau}\,\bar{L}_\mu H \tau_R \\[2mm]
&+ y_{\tau e}\,\bar{L}_\tau H e_R
+ y_{\tau\mu}\,\bar{L}_\tau H \mu_R
+ y_{\tau\tau}\,\bar{L}_\tau H \tau_R \\[3mm]
&+ y_{e1}\,\bar{L}_e  H \Sigma_1
+ y_{e2}\,\bar{L}_e  H \Sigma_2 
+ y_{e3}\,\bar{L}_e  H \Sigma_3\\[2mm]
&+ y_{\mu1}\,\bar{L}_\mu H \Sigma_1 
+ y_{\mu2}\,\bar{L}_\mu  H \Sigma_2 
+ y_{\mu3}\,\bar{L}_\mu  H \Sigma_3\\[2mm]
&+ y_{\tau1}\,\bar{L}_\tau H \Sigma_1 
+ y_{\tau2}\,\bar{L}_\tau  H \Sigma_2 
+ y_{\tau3}\,\bar{L}_\tau  H \Sigma_3\\[2mm]
&+ M_{11}\,\bar{\Sigma}_1^c \Sigma_1
+ M_{22}\,\bar{\Sigma}_2^c \Sigma_2 
+  M_{33}\,\bar{\Sigma}_3^c \Sigma_3  \\[2mm]
& +M_{12}\big(\bar{\Sigma}_1^c \Sigma_2+\bar{\Sigma}_2^c \Sigma_1\big)
 +M_{13}\big(\bar{\Sigma}_1^c \Sigma_3+\bar{\Sigma}_3^c \Sigma_1\big)
 +M_{23}\big(\bar{\Sigma}_2^c \Sigma_3+\bar{\Sigma}_3^c \Sigma_2\big) \\[2mm]
& + y_{ii}\,\bar{\Sigma}_i^c \Sigma_i \chi +  y_{ii}\,\bar{\Sigma}_i^c \Sigma_i \chi* +  y_{ij}\,(\bar{\Sigma}_i^c \Sigma_j + \bar{\Sigma}_j^c \Sigma_i) \chi +   y_{ij}\,(\bar{\Sigma}_i^c \Sigma_j + \bar{\Sigma}_j^c \Sigma_i) \chi^*.
\end{split}
\label{lagrangian}
\end{equation}
\noindent The Yukawa interactions involving $\chi$ and the fermion triplets explicitly break the $\mathbb{Z}_3$ symmetry, inducing field-dependent fermion masses $i.e.$ $m_i(\chi)=M_i+y_i\chi$ that generate a radiative vacuum energy bias through quantum corrections. The resulting lifting of the vacuum degeneracy renders the domain-walls metastable, leading to their eventual annihilation and the production of a stochastic gravitational-wave background.

\noindent From the Lagrangian given in Eqn. (\ref{lagrangian}), the charged lepton mass matrix is given as
\begin{equation}
  M_l = \frac{v_H}{\sqrt{2}}\begin{pmatrix}
      y_{ee} & y_{e\mu} & y_{e\tau}\\
      y_{\mu e} & y_{\mu \mu} & y_{\mu \tau}  \\
      y_{\tau e} & y_{\tau \mu} & y_{\tau \tau} 
  \end{pmatrix}  .
\end{equation}
Here $v_H$ is the vacuum expectation value ($vev$) of the Higgs field. The Dirac mass matrix and the fermion triplet mass matrix is given as 
\begin{equation}
  M_D = \frac{v_H}{\sqrt{2}}\begin{pmatrix}
      y_{e1} & y_{e2} & y_{e3}  \\
      y_{\mu 1} & y_{\mu 2} & y_{\mu 3}  \\
      y_{\tau 1} & y_{\tau 2} & y_{\tau 3} 
  \end{pmatrix},  
\quad
  M_\Sigma = \begin{pmatrix}
      M_1 & M_{21}  & M_{31}  \\
     M_{21}   & M_2  & M_{23} \, e^{i \delta} \\
     M_{31} & M_{23}\, e^{i \delta} & M_{3}  
  \end{pmatrix}. 
\end{equation}
\noindent The fermion triplet mass matrix ($M_\Sigma$) can be diagonalized as $U_R M_\Sigma U_R^T = \text{Diag}(M_{\Sigma_1},M_{\Sigma_2}, M_{\Sigma_3})$. The active neutrino mass matrix is obtained by the Type-III seesaw mechanism is given by
\begin{equation}
M_\nu = -M_D M_\Sigma^{-1} M_D^T.
\label{neutrinomassmatrix}
\end{equation}
This neutrino mass matrix is diagonalized by a unitary mixing matrix to yield predictions for neutrino observables. It is to be noted that the presence of only a single discrete symmetry in our framework introduces a large number of free parameters. To make the model more predictive, we impose a set of simplifying assumptions. In the charged lepton mass matrix ($M_l$), we assume symmetric Yukawa couplings such that 
$y_{e\mu} = y_{\mu e}$ = $y_{e\tau} = y_{\tau e}$ = $y_{\tau\mu} = y_{\mu\tau}$. Similarly, in the Dirac neutrino mass matrix, we impose the relations 
$y_{e2} = y_{\mu 1}$ = $y_{e3} = y_{\tau 1}$ = $y_{\tau 2} = y_{\mu 3}$. For the Majorana mass matrix, we consider degenerate diagonal entries 
$M_1 = M_2 = M_3$, along with equal off-diagonal terms 
$M_{21} = M_{31} = M_{23}$. These assumptions significantly reduce the number of independent parameters in the model. We then perform a $\chi^2$ analysis to test the consistency of the model with experimental data.
In the following section, we will numerically diagonalize the neutrino mass matrix extracted in Eqn. (\ref{neutrinomassmatrix}) and elucidate predictions for the neutrino observables.

\section{Neutrino Mass and Mixing}\label{section3}
\noindent The neutrino mass matrix given by Eqn. (\ref{neutrinomassmatrix}) is diagonalized using the relation $U_\nu^T M_\nu U_\nu = \text{diag}(m_{\nu_1}, m_{\nu_2}, m_{\nu_3})$. Now, the mixing angle can be extracted from $U_{\text{PMNS}}$ as
\begin{equation}
    \sin^2 \theta_{13} = |U_{13}|^2, \quad
    \sin^2 \theta_{12} = \frac{|U_{12}|^2}{1 - |U_{13}|^2}, \quad
    \sin^2 \theta_{23} = \frac{|U_{23}|^2}{1 - |U_{13}|^2}.
\end{equation}
The Dirac $CP$-violating phase($\delta_{CP}$) can be determined from the PMNS matrix elements through the Jarlskog invariant, defined as
\begin{equation}
    J_{CP} = \text{Im}\left[U_{11} U_{22} U^{*}_{12} U^{*}_{21}\right] = s_{23} c_{23} s_{12} c_{12} s_{13} c_{13}^2 \sin\delta_{CP},
\end{equation}
where \( s_{ij} = \sin\theta_{ij} \) and \( c_{ij} = \cos\theta_{ij} \). In addition to $\delta_{CP}$, the Majorana $CP$ phases can be investigated using the PMNS matrix elements as
\begin{equation}
    I_1 = \text{Im}\left[U_{11}^* U_{12}\right] = c_{12} s_{12} c_{13}^2 \sin\left(\frac{\alpha_{21}}{2}\right),
\end{equation}
\begin{equation}
    I_2 = \text{Im}\left[U_{11}^* U_{13}\right] = c_{12} s_{13} c_{13} \sin\left(\frac{\alpha_{31}}{2} - \delta_{CP}\right).
\end{equation}
\noindent The effective Majorana mass($m_{\beta \beta}$) is given as
\begin{equation}
m_{\beta\beta}
= \left| 
m_1 U_{e1}^2
+ m_2 U_{e2}^2 e^{i\alpha_{21}}
+ m_3 U_{e3}^2 e^{i\alpha_{31}}
\right| .
\end{equation}

\begin{figure}
    \centering
    \includegraphics[width=1\linewidth]{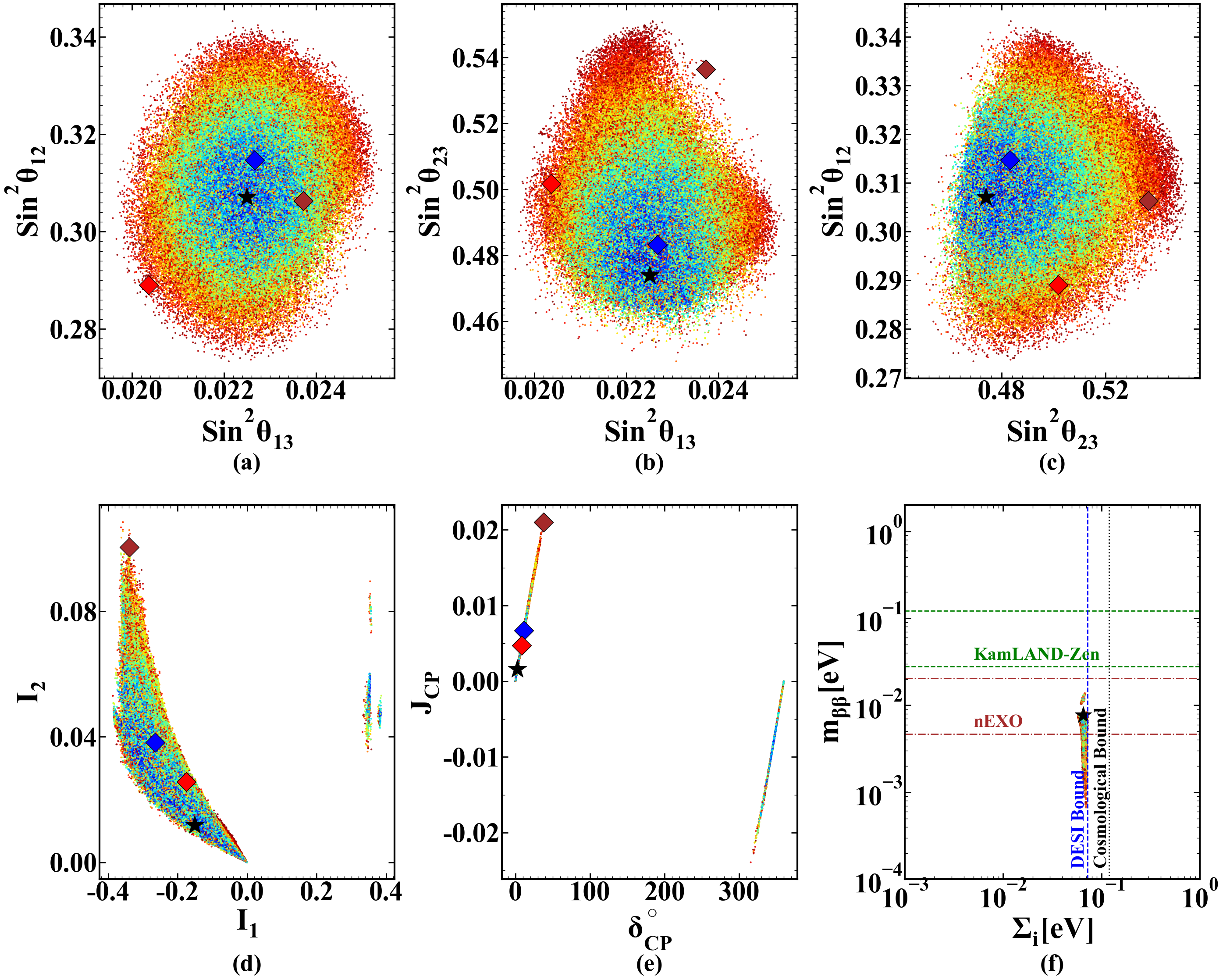}
    \includegraphics[width =0.5\linewidth]{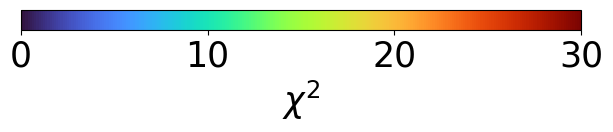}
    \caption{Correlation among the neutrino mixing angles, CP invariants($I_1$ and $I_2$), Jarlskog invariant($J_{CP}$) and Dirac Type CP phase ($\delta_{CP}$) and effective Majorana mass($m_{\beta \beta}$) and sum of neutrino masses($\Sigma_i$). The color bar represents the $\chi^2$ values ranging from 0 to 30. The `$*$' symbol indicates the best-fit point corresponding to the minimum chi-square value, $\chi^2_{\min} = 0.24$. The blue diamond represents the selected Benchmark Point 2 (BP2), the red diamond denotes Benchmark Point 3 (BP3) and the brown diamond denotes Benchmark Point 4 (BP4).}
    \label{mixing_angle}
\end{figure}

\begin{table*}[t]
\small
    \centering
    \renewcommand{\arraystretch}{1} 
    \resizebox{\textwidth}{!}{\begin{tabular}{l l l l l} 
        \hline \hline
        Parameter & best-fit$\pm1\sigma$ range (NH) & best-fit$\pm1\sigma$ range (IH) & $3\sigma$ range (NH) & $3\sigma$ range (IH) \\
        \hline
        $\sin^2\theta_{12}$ & $0.308^{+0.0067}_{-0.0066}$ & $0.308^{+0.0067}_{-0.0066}$ & $0.289 - 0.329$ & $0.289 - 0.329$ \\
        
        $\sin^2\theta_{23}$ & $0.470^{+0.017}_{-0.014}$ & $0.550^{+0.013}_{-0.016}$ & $0.435 - 0.584$ & $0.439 - 0.584$ \\
        
        $\sin^2\theta_{13}$ & $0.02248^{+0.00055}_{-0.00059}$ & $0.02262^{+0.00057}_{-0.00056}$ & $0.02064 - 0.02418$ & $0.02093 - 0.02441$ \\
        
        $\Delta m^2_{3l} \times 10^{-3} \text{eV}^2$ & $2.511^{+0.021}_{-0.020}$ & $-2.483^{+0.020}_{-0.020}$ & $2.450 - 2.576$ & $-2.547 - -2.421$ \\
        
        $\Delta m^2_{21} \times 10^{-5} \text{eV}^2$ & $7.53^{+0.094}_{-0.10}$ & $7.53^{+0.094}_{-0.10}$ & $7.26 - 7.82$ & $7.23 - 7.82$ \\
        
        $m_e/m_\mu$ & $0.004737$ & $0.004737$ & -- & -- \\
        
        $m_\mu/m_\tau$ & $0.058823 $ & $0.058823$ & -- & -- \\
        \hline \hline
    \end{tabular}}
    
    \caption{The neutrino oscillation data used in the numerical analysis taken from NuFIT 6.1. The central values of the charged lepton mass ratios are taken from~\cite{Esteban:2024eli, ParticleDataGroup:2024cfk,NuFit61}. During the scan of the model parameter space, the uncertainties in these ratios as fixed at 0.1\% of their respective central values~\cite{Xing:2007fb}.}
    \label{tab2}
\end{table*}

\noindent In this work, we have investigated the Type-III seesaw mechanism extended by a complex scalar within the framework of $\mathbb{Z}_3$ symmetry. To examine the consistency of the model with current experimental data, we have performed a chi-square analysis and identified a $\chi^2_{\min} = 0.24$ corresponding to the best-fit region of the parameter space. The chi-square function used in the analysis is defined as
\begin{equation}
\chi^2 = \sum_{i=1}^{7} \left( \frac{P_i - P_i^{0}}{\sigma_i} \right)^2,
\label{chisq}
\end{equation}
where, $P_i$ is the prediction of the neutrino observable from the model, $P_i^0$ is the best fit value from NuFIT 6.1 summarized in Table~\ref{tab2}, and $\sigma_i$ shows the uncertainty of neutrino observables at $1\sigma$ level~\cite{Esteban:2024eli, NuFit61,ParticleDataGroup:2024cfk}. In this work, we have sampled seven neutrino observables, i.e, three mixing angles($\sin^2\theta_{13}$, $\sin^2\theta_{12}$, $\sin^2\theta_{23}$), two mass squared differences ($\Delta m_{21}^2$ and $\Delta m_{3l}^2$) and two charged lepton mass ratios($\frac{m_e}{m_\mu}$ and $\frac{m_\mu}{m_\tau}$). The Dirac CP phase is not included as an input parameter due to its relatively weak experimental constraint at present. Fig.~\ref{mixing_angle}(a),(b) and (c) shows that mixing angles are compatible with the current experimental data at 3$\sigma$ level. The atmospheric mixing angle $\sin^2\theta_{23}$ is predicted to lie in first octant. This prediction is testable at long baseline(LBL) experiments such as No$\nu$a~\cite{2852068}, DUNE~\cite{DUNE:2020fgq}, Hyper-Kamiokande~\cite{Hyper-Kamiokande:2018ofw}. Fig.~\ref{mixing_angle}(d) shows the correlation between CP invariants ($I_1$ and $I_2$). Fig.~\ref{mixing_angle}(e) presents the correlation between Jarlskog invariant($J_{CP}$) and $\delta_{CP}$. The $\delta_{CP}$ lies in the first and fourth octant, these predictions will be testable at LBL experiments~\cite{2852068, DUNE:2020fgq, Hyper-Kamiokande:2018ofw}. The correlation between effective Majorana mass($m_{\beta \beta}$) with sum of neutrino masses($\Sigma_i$) is shown in Fig.~\ref{mixing_angle}(f). The $m_{\beta \beta} $ is compatible with nEXO experiment~\cite{nEXO:2021ujk}. The sum of neutrino masses is compatible with the cosmological bound ($\Sigma_i <0.12 \text{eV}$ at 95\% confidence level(CL)~\cite{Planck:2018vyg} as well as the stringent bound proposed by DESI experiment ($\Sigma_i < 0.072 \text{eV}$ at 95\% CL)~\cite{DESI:2024mwx}. The sum of neutrino mass is constrained in the region $(0.056-0.082) \text{eV}$. Future results from DESI experiment will serve as an important test of the predictability of the model. In the next section, we will explore baryogenesis through leptogenesis.

\section{Baryogenesis from the Leptogenesis}\label{section4}
\noindent Leptogenesis provides a compelling framework for addressing one of the most fundamental questions in particle physics and cosmology: the origin of the observed matter–antimatter asymmetry of the Universe~\cite{Fukugita:1986hr, Covi:1996wh,Buchmuller:2005eh,Buchmuller:2004nz}. The generation of baryon asymmetry through leptogenesis requires the fulfillment of the three Sakharov conditions, namely baryon number violation, violation of C and CP symmetries, and a departure from thermal equilibrium during the cosmological evolution of the Universe~\cite{Sakharov:1967dj}. Although the SM formally accommodates these conditions in an expanding Universe, the magnitude of CP violation predicted within the SM is insufficient to reproduce the observed baryon asymmetry. Consequently, the existence of additional sources of CP violation becomes indispensable. In this context, the lepton sector emerges as a well-motivated candidate for new CP-violating effects, although such sources have not yet been conclusively established experimentally~\cite{Kuzmin:1985mm}. In this framework, we briefly discuss the baryogenesis $\textit{via}$ leptogenesis through the decay of the lightest SU(2) fermion triplet, which generates CP asymmetry, which has been thoroughly discussed in the literature~\cite{AristizabalSierra:2010mv,Mishra:2022egy,Vatsyayan:2022rth,Chaudhuri:2026dvc,Singh:2023eye,Mahapatra:2023dbr,Mishra:2020gxg,Abhishek:2026hex,Baldes:2025uwz,Suematsu:2019kst,Datta:2021elq,Priya:2026ehe}. This asymmetry subsequently contributes to the generation of net $B-L$ asymmetry in the early Universe. For thermal leptogenesis, there exists a lower bound on the mass of the lightest decaying fermion, commonly known as the Davidson-Ibarra bound~\cite{Davidson:2002qv,Davidson:2008bu}. In the case of Type-III seesaw, however, the presence of electroweak gauge interactions involving the $SU(2)$ fermion triplets significantly enhances the washout processes, thereby modifying the lower bound on the triplet mass scale to approximately $3 \times 10^{10}~\mathrm{GeV}$~\cite{Hambye:2003rt, Vatsyayan:2022rth}. After the spontaneous breaking of the $\mathbb{Z}_3$ symmetry, the scalar singlet $\chi$ acquires a non-zero vacuum expectation value. As a consequence, the Yukawa interactions between $\chi$ and the fermion triplets generate field-dependent fermion masses of the form
\begin{eqnarray}
m_i(\chi)=M_i+y_i\chi \hspace{0.4cm} \text{with} \hspace{0.4cm} i = 1,2,3.
\end{eqnarray}
These field-dependent masses play an important role in leptogenesis and also contribute to the radiative lifting of the vacuum degeneracy through the Coleman-Weinberg effective potential, which will be discussed later in the domain-wall section. The CP asymmetry parameter associated with the decay of the fermion triplet is given by
\begin{equation}
    \epsilon_i = -\sum_{j = 2,3} \frac{3}{2}\frac{m_{i}}{m_{j}} \frac{\Gamma_j}{m_{j}} I_j \frac{V_j - 2 S_j}{3},
    \end{equation}
\noindent where
\begin{equation}
    I_j = \frac{Im[(\tilde{Y}^\dagger \tilde{Y})_{ij}^2]}{(\tilde{Y}^\dagger \tilde{Y})_{ii}(\tilde{Y}^\dagger \tilde{Y})_{jj}}.
\end{equation}
Here,  $\tilde{Y} = y_d U_L U_R$, with $y_d = \frac{M_D}{v_H}$, while $V_j$ and $S_j$ are the loop factors associated with the vertex and self-energy corrections, respectively, given by 
\begin{equation}
    V_j = \frac{m_{j}^2 (m_{j}^2 - m_{i}^2)}
{(m_{j}^2 - m_{i}^2)^2 + m_{i}^2 \Gamma_{\Sigma_j}^2},
\end{equation}
\quad
\begin{equation}
    S_j = 2 \frac{m_{j}^2} {m_{i}^2} \left( \left( 1 + \frac{m_{j}^2}{m_{i}^2} \right)  
\ln \left( 1 + \frac{m_{j}^2}{m_{i}^2} \right) - 1 \right).
\end{equation}
The numerical ranges of coupling $y_d$ consistent with the neutrino oscillation data have been given in Appendix \ref{appendix1}. In the hierarchical limits($m_{1}< m_{2} < m_{3} $), the loop factors reduce to unity \cite{Hambye:2012fh,Albright:2003xb,Mishra:2022egy}. Further, $\Gamma_j$ represents the decay width of the triplet fermion and can be expressed as 
\begin{equation}
\Gamma_{\Sigma_j} = \left( \frac{|(\tilde{Y}^\dagger \tilde{Y})_{jj}|}{8\pi} \right) m_{j}.
\end{equation}
\noindent The Boltzmann equation (BEs) plays an important role in tracking the evolution of lepton asymmetry as the universe gradually cools down over time. The relevant coupled BEs are given by
\begin{equation}
s H_o z \frac{dY_\Sigma}{dz} = -\gamma_D \left( \frac{Y_\Sigma}{Y^{\text{eq}}_\Sigma} - 1 \right) - 2 \gamma_A \left( \frac{Y_\Sigma^2}{(Y^{\text{eq}}_\Sigma)^2} - 1 \right),
\label{ysigma}
\end{equation}
\begin{equation}
s H_o z \frac{dY_{B-L}}{dz} = -\gamma_D \, \epsilon_\Sigma \left( \frac{Y_\Sigma}{Y^{\text{eq}}_\Sigma} - 1 \right) - \frac{Y_{B-L}}{Y^{\text{eq}}_l} \left( \frac{\gamma_D}{2} + \gamma^{\text{sub}}_\Sigma \right),
\label{ybl}
\end{equation}
where, $Y_{\Sigma a}$ = $\frac{n_a(z)}{s(z)}$ represents the number density of particle species $a$. $s(z)$ is the entropy density given as $s(z)= 0.44 g_* T^3$, $z$ is a dimensionless parameter varying inversely with the temperature of the universe, given as $z=m_{1}/T$. $H_o$ represents the Hubble expansion rate, given as $H_o = 1.66 \, \frac{\sqrt{g_*} \, T^2}{M_{\text{Pl}}}$. $\gamma$ denotes the reaction density of processes under consideration, and `D' denotes the decay processes, given as 
\begin{equation}
    \gamma _D (z) = s(z) Y_\Sigma^{eq} \Gamma _{\Sigma} \frac{K_1(z)}{K_2(z)},
\end{equation}
\noindent where $K_1(z)$ and $K_2(z)$ are the modified Bessel functions and $\gamma _A$ in Eqn. (\ref{ysigma}) denotes the gauge annihilation processes and represented as 
\begin{equation}
\gamma _A(z) = \frac{m_{1} T^3}{32\pi^3} e^{-2z}  
\left[
\frac{111 g^4}{8\pi} + \frac{3}{2z} \left( \frac{111 g^4}{8\pi} + \frac{51 g^4}{16\pi} \right) + \mathcal{O}\left(\frac{1}{z^2}\right)
\right].
\end{equation}
\noindent Here, $g$ is the typical gauge coupling. The $\gamma^{\text{sub}}$ is for washout effects by $\Delta$L = 2 processes. The equilibrium yields are given by
\begin{equation}
    Y_\Sigma^{eq} = \frac{135 g_\Sigma}{16 \pi^4 g_*} z^2 K_2(z), \quad Y_l^{eq} = \frac{3}{4} \frac{45 \zeta(3) g_l}{2 \pi^4 g_*},
\end{equation}
\noindent where $g_l = 2$, $g_\Sigma = 2$ and $g_* = 106.75$. For a lepton asymmetry to survive, the decays of heavy fermion triplets must occur out of thermal equilibrium. This condition is satisfied when the decay rate of the heavy fermions is not significantly larger than the Hubble expansion rate of the Universe at a temperature equal to their mass, $T = m_{1}$. Under such circumstances, the generated lepton asymmetry is not completely washed out. The departure from thermal equilibrium is commonly quantified by the decay parameter $k$, defined as the ratio of the decay width of the heavy fermion to the Hubble expansion rate evaluated at $T = m_{1}$,
\begin{eqnarray}
k = \frac{\Gamma_\Sigma}{H_o(T = m_{1})}
    = \frac{\tilde{m}_i}{m^*} \hspace{0.5cm} \text{with} \hspace{0.5cm} \tilde{m}_i = \frac{(\tilde{Y}^\dagger \tilde{Y})_{ii} v^2}{m_{i}},
\end{eqnarray}
while $m^*$ denotes the equilibrium neutrino mass scale evaluated at $T = m_{1}$, the washout parameter $k$ allows one to distinguish three different regimes: $k \ll 1$ corresponds to the weak washout regime, $k \sim 1$ to the intermediate washout regime, and $k \gg 1$ to the strong washout regime~\cite{Marciano:2024nwm,Davidson:2008bu,Priya:2025wdm}. For all the selected benchmark points, we find that the framework lies in the strong washout regime. In this regime, the washout dynamics are predominantly governed by inverse decay processes, whereas the contributions from $\Delta L = 1$ and $\Delta L = 2$ scattering processes remain subdominant and can therefore be neglected to a good approximation. Consequently, following Refs.~\cite{Marciano:2024nwm,Davidson:2008bu,Priya:2025wdm}, our analysis includes only the decay processes $\Sigma_1 \rightarrow LH$ (and its CP-conjugate channel) together with the corresponding inverse decay processes $LH \rightarrow \Sigma_1$.
\begin{table}[]
    \centering
    \begin{tabular}{c c c c c}
    \hline \hline
      Benchmark Point &  $\chi_{min}^2$ &$m_{1}$(GeV) & $\epsilon_{CP}$ & $v_\chi$ \text{(GeV)}\\
      \hline
       BP1 & $0.24$ & $1.05 \times 10^{9}$ & $-3.48 \times 10^{-2}$ & $5.55\times10^{11}$\\
       BP2 & $14.99$ & $1.53\times 10^{9}$ & $-4.89 \times 10^{-3}$ & $5.93 \times10^{11}$ \\
       BP3 & $29.99$ & $5.36\times 10^{9}$ & $-4.54 \times 10^{-4}$ & $1.38 \times10^{12}$\\
       BP4 & $29.17$ & $4.51\times 10^{9}$ & $ -1.19 \times 10^{-3}$ & $5.93 \times10^{11}$\\
       \hline \hline
    \end{tabular}
    \caption{List of benchmark points used in the analysis of thermal leptogenesis and the resulting gravitational-wave spectra. All benchmark points are consistent with the current neutrino oscillation data within the Type-III seesaw framework.}
\label{tab:BPs}
\end{table}

\begin{figure}
    \centering
    \includegraphics[width=0.49\linewidth]{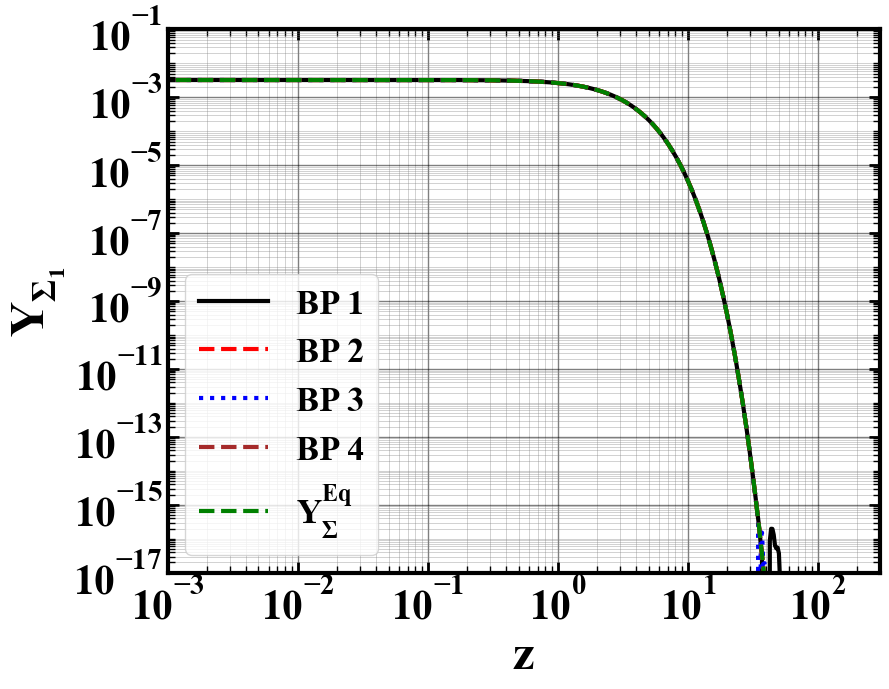}
    \includegraphics[width=0.49\linewidth]{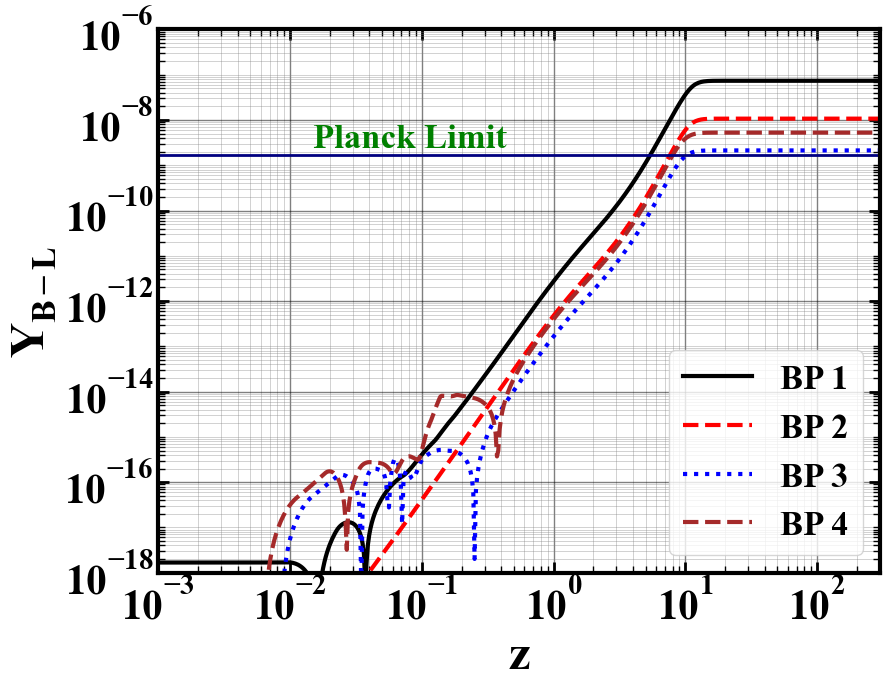}
    \caption{Evolution of the comoving number density of the lightest fermion triplet $Y_{\Sigma_1}$ (\textbf{left}) and the $B-L$ asymmetry $Y_{B-L}$ (\textbf{right}) as functions of $z = m_{1}/T$ for four benchmark points (BP1, BP2, BP3, BP4) given in Table \ref{tab:BPs}. The required value of lepton asymmetry from the Planck cosmological date is depicted by the horizontal line in the right plot~\cite{Planck:2018vyg}.}
    \label{lepto}
\end{figure}

\noindent The results of the numerical analysis have been presented in Fig. \ref{lepto}. We solve the two coupled Boltzmann equations given in Eqn. \ref{ysigma} and \ref{ybl} for four BPs given in Table \ref{tab:BPs}. The left panel of Fig. \ref{lepto} shows the evolution of the comoving number density of the fermionic triplet, $Y_{\Sigma}$, as a function of $z=m_{1}/T$ for the four benchmark points listed in Table~\ref{tab:BPs}. At very small values of $z$ ($T \gg m_{1}$), the triplet abundance closely follows its equilibrium distribution, indicating that the gauge-mediated interactions are sufficiently strong to maintain thermal equilibrium in the early Universe. As the temperature decreases and $z \gtrsim \mathcal{O}(1)$, the equilibrium abundance becomes Boltzmann suppressed and $Y_{\Sigma}$ starts deviating from the equilibrium curve. Eventually, for $z \sim 10$-$30$, the triplet abundance rapidly drops due to the decay of the heavy fermionic triplets into lepton and Higgs doublets, signaling the freeze-out and decay epoch relevant for leptogenesis. One can also observe that all four benchmark points exhibit a very similar thermal history for the triplet abundance, implying that the dominant differences in the generated baryon asymmetry arise mainly from the CP asymmetry parameter $\epsilon_{\rm CP}$ and the washout strength characterized by the effective neutrino mass parameter. The right panel of Fig.~\ref{lepto}, we show the evolution of the normalized $B-L$ asymmetry, $Y_{B-L}$, as a function of $z=\frac{m_{1}}{T},$ for the four benchmark points listed in Table~\ref{tab:BPs}. At early times ($z\ll 1$), the fermionic triplets remain in thermal equilibrium due to rapid gauge scatterings and inverse decay processes, resulting in a highly suppressed asymmetry. As the temperature drops below the triplet mass scale ($z\gtrsim 1$), the reaction rates gradually become smaller than the Hubble expansion rate and the triplet decays occur out of equilibrium. In this regime, the CP-violating decays of $\Sigma_1$ generate a net lepton asymmetry, which subsequently grows rapidly with increasing $z$. The asymmetry eventually reaches a plateau around $z\sim \mathcal{O}(10)$, indicating the freeze-out of washout processes and the conservation of the generated $B-L$ asymmetry. The magnitude of the final asymmetry is governed by the interplay between the CP asymmetry parameter $\epsilon_{\rm CP}$, the triplet mass $m_1$, and the symmetry-breaking scale $v_\chi$, which determine both the source and washout terms in the Boltzmann equations. In particular, BP1, characterized by a relatively larger effective CP asymmetry and comparatively weaker washout effects, yields the largest final asymmetry, whereas BP3 exhibits the smallest asymmetry due to stronger washout suppression. For all benchmark points, the final asymmetry lies in the range $Y_{B-L}\sim 10^{-9}-10^{-8},$ which, after sphaleron conversion, is compatible with the observed baryon asymmetry of the Universe inferred from Planck measurements. These results demonstrate that the proposed Type-III seesaw framework naturally accommodates successful thermal leptogenesis for triplet masses of order $m_{1}\sim 10^{9}~{\rm GeV}$. In the following section, we investigate the domain-wall phenomenology of the model and establish its connection to the leptogenesis sector, with particular emphasis on the dependence of the domain-wall dynamics and gravitational-wave signatures on the heavy fermion mass parameters $m_i$ entering the leptogenesis framework.
\section{domain-walls induced Gravitational Waves}\label{section5}
\noindent The spontaneous breaking of symmetries in the early Universe can give rise to primordial phase transitions, which may be accompanied by the formation of topological defects~\cite{Saikawa:2017hiv,Roshan:2026yon,Vilenkin:1984ib,Kibble:1976sj,Zeldovich:1974uw}. In particular, the spontaneous breaking of a discrete symmetry leads to the formation of two-dimensional, surface-like defects known as domain-walls. These structures separate spatial regions corresponding to different degenerate vacua of the theory and are typically formed during the early stages of cosmological evolution. Once formed, domain-walls rapidly evolve toward a scaling regime, in which the characteristic number of domain-walls per Hubble volume remains approximately constant within a few Hubble times. As a consequence, the energy density stored in domain-walls decreases more slowly with the expansion of the Universe than that of radiation or matter. Specifically, while radiation and matter scale as $a^{-4}$ and $a^{-3}$, respectively, the energy density of domain-walls scales as $a^{-1}$, where $a(t)$ denotes the cosmological scale factor. As a result, if stable, domain-walls would eventually dominate the total energy density of the Universe, leading to severe cosmological consequences, including distortions of large-scale structure formation and conflicts with primordial nucleosynthesis constraints.

\noindent The existence of stable domain-walls associated with the spontaneous breaking of a discrete symmetry, such as a $\mathbb{Z}_3$ symmetry, is severely constrained by cosmological observations. In particular, domain-walls with a symmetry-breaking scale as low as $\mathcal{O}(1)\,\mathrm{MeV}$ are excluded, since their persistence would lead to unacceptably large anisotropies in the cosmic microwave background (CMB) observed today \cite{Hiramatsu:2010yz}. This follows from the fact that stable domain-walls eventually dominate the energy density of the Universe, thereby spoiling the standard cosmological evolution. Nevertheless, it remains possible to consider unstable domain-walls that decay sufficiently early, thus avoiding conflicts with cosmological observations. One well-motivated mechanism to render domain-walls unstable is to assume that the discrete symmetry is only approximate, rather than exact. In this case, a small explicit breaking term introduces a bias in the scalar potential, lifting the degeneracy among the three vacua and generating a pressure that drives the collapse and decay of the domain-walls. Such explicit breaking effects can naturally arise from higher-dimensional, non-renormalizable operators suppressed by a cutoff scale, for instance the Planck scale. The idea of resolving the domain-wall problem through an approximate discrete symmetry has been extensively studied in the literature~\cite{Vilenkin:1984ib, Hiramatsu:2010yz,Hiramatsu:2013qaa, Kitajima:2015nla, Saikawa:2017hiv, Krajewski:2017czs, Chen:2020wvu, Borah:2026kfo, Ma:2025bjf, Roshan:2026yon,Gouttenoire:2025ofv, Wu:2022stu,Barreto:2026igt, Li:2025gld, Fu:2024jhu, Wu:2022tpe, Wei:2022poh,Bhandari:2026ujy,Borah:2025bfa, Barman:2022yos}. An alternative possibility for generating unstable domain-walls involves a post-inflationary, non-thermal phase transition, in which the discrete symmetry is broken after inflation, leading to the formation of domain-walls that decay before they come to dominate the energy density of the Universe.
 In this work, we study the production of gravitational waves from the annihilation of unstable domain-walls. The renormalizable $\mathbb{Z}_3$ invariant potential of $\chi$ is given as
\begin{equation}
    V(\chi) = -\mu^2\chi \chi^* + \lambda_1 (\chi \chi^*)^2 + \lambda_2 (\chi^3 + \chi^{*3}).
\label{potential}
\end{equation}
\noindent The complex scalar field ($\chi$) can be written as
\begin{equation}
\chi = \rho e^{i\theta}, \qquad \chi^\ast = \rho e^{-i\theta},
\end{equation}
here $\rho$ is the magnitude and $\theta $ is the phase of complex field $\chi$. Hence the potential is modified as
\begin{equation}
V(\rho,\theta) = -\mu^{2}\rho^{2} + \lambda_{1}\rho^{4} -2\lambda_{2}\mu\rho^{3}\cos(3\theta).
\end{equation}
\noindent Minimization with respect to $\theta$
\begin{equation}
\frac{\partial V}{\partial \theta} = 6 \lambda_2 \mu \rho^3 \sin(3\theta) = 0.
\end{equation}
and
\begin{equation}
\sin(3\theta)=0 \Rightarrow 3\theta = 2\pi k \Rightarrow \theta = \frac{2\pi k}{3}, \quad k=0,1,2.
\end{equation}
\noindent To minimize the potential V, we set $\cos(3\theta)=1$ in $V(\rho,\theta)$. Minimization with respect to $\rho$ can be proceed as 
\begin{equation}
\frac{\partial V}{\partial \rho} = -2\mu^2 \rho + 4\lambda_1 \rho^3 - 6\lambda_2 \mu \rho^2 = 0.
\end{equation}

\noindent Divide by $2\rho$ (for $\rho>0$)
\begin{equation}
-\mu^2 + 2\lambda_1 \rho^2 - 3\lambda_2 \mu \rho = 0.
\end{equation}

\begin{equation}
2\lambda_1 \rho^2 - 3\lambda_2 \mu \rho - \mu^2 = 0.
\end{equation}

\noindent After solving the quadratic equation, we get
\begin{equation}
\rho = \frac{3\lambda_2 \mu \pm \sqrt{9\lambda_2^2 \mu^2 + 8\lambda_1 \mu^2}}{4\lambda_1}.
\end{equation}

\noindent The negative root is discarded because $\rho \ge 0$. Define $\beta = \frac{3\lambda_2}{\sqrt{8\lambda_1}}$, then the physical solution becomes
\begin{equation}
\rho = \frac{\mu}{\sqrt{2\lambda_1}}\left(\beta + \sqrt{1+\beta^2}\right).
\end{equation}

\noindent After reinserting the phase, finally, we have 
\begin{equation}
\langle \chi \rangle = v_k = \frac{\mu}{\sqrt{2\lambda_1}}\left(\beta + \sqrt{1+\beta^2}\right) e^{i\frac{2\pi k}{3}}, \qquad k = 0,1,2.
\end{equation}
\noindent Here, all the coefficients are real and positive. The Yukawa interaction between $\chi$ and the fermion triplets explicitly breaks the $\mathbb{Z}_3$ symmetry, which plays a crucial role in facilitating domain-wall annihilation. The last line of the Lagrangian in Eqn.~\ref{lagrangian} contains the terms responsible for this non-invariance. If stable, such domain-walls would rapidly come to dominate the energy density of the Universe, leading to conflicts with constraints from the cosmic microwave background (CMB) and big bang nucleosynthesis (BBN). Assuming that these domain-walls form after inflation, they can be removed by introducing a small pressure difference across them, commonly referred to as a bias term, which explicitly breaks the $\mathbb{Z}_3$ symmetry. 

\noindent In many scenarios, as discussed in the first section, such bias terms are introduced phenomenologically as operators involving non-trivial $\mathbb{Z}_3$ charge combinations of the scalar field that explicitly break the $\mathbb{Z}_3$ symmetry, without a direct connection to other sectors of the theory. In this work, however, we demonstrate that the required bias term is generated radiatively through Yukawa interactions involving the same heavy fermionic states responsible for neutrino mass generation $\textit{via}$ the Type-III seesaw mechanism, where neutrinos acquire a Majorana nature.
We introduce a Yukawa interaction between the $\mathbb{Z}_3$-charged scalar $\chi$ and the fermionic triplets $\Sigma$, which explicitly breaks the $\mathbb{Z}_3$ symmetry. At tree level, the scalar potential preserves the discrete symmetry, leading to multiple degenerate minima related by $\mathbb{Z}_3$ transformations. These domain-walls are characterized by their surface energy density, commonly referred to as the domain-wall tension, which depends on the parameters of the scalar potential governing the $\mathbb{Z}_3$-charged field. During the cosmological phase transition, the scalar field $\chi$ acquires a non-zero vacuum expectation value, leading to the spontaneous breaking of the $\mathbb{Z}_3$ symmetry. The vacuum structure consists of three degenerate minima,
\begin{equation}
\langle \chi \rangle = v_\chi \, \omega^k, \quad k = 0,1,2, \quad \omega = e^{2\pi i/3}.
\end{equation}
In the presence of Yukawa interactions, the fermion masses become field-dependent,
\begin{equation}
m_i(\chi) = M_i + y_i \chi,
\end{equation}
and hence take different values at each vacuum,
\begin{equation}
m_i^{(k)} = M_i + y_i v_\chi \omega^k.
\end{equation}
\noindent For $\mathbb{Z}_3$ symmetry, the vacuum structure consists of three minima related by non-trivial complex phases. If all couplings are taken to be real, two of these vacua can remain physically indistinguishable, leading to a residual degeneracy. To avoid this, we consider complex Yukawa couplings between the fermion triplets and the complex scalar $\chi$. Consequently, the field-dependent fermion masses acquire distinct values in the three vacua, allowing the vacuum degeneracy to be completely lifted through radiative corrections. At the quantum level, fermions running in the loop generate a Coleman--Weinberg correction to the scalar potential~\cite{Coleman:1973jx},
\begin{equation}
V_{\text{CW}}(\chi) = \sum_i \frac{n_i}{64\pi^2} \, m_i^4(\chi) 
\left[ \ln\left(\frac{m_i^2(\chi)}{\mu^2}\right) - \frac{3}{2} \right],
\end{equation}
where $n_i$ denotes the degrees of freedom and $\mu$ is the renormalisation scale. Due to the explicit breaking of $\mathbb{Z}_3$ through the Yukawa interaction, the effective potential evaluated at the different vacua is no longer degenerate. This generates a non-zero energy difference between the minima,
\begin{equation}
\Delta V_{ij} = V_{\text{CW}}(v_\chi \omega^i) - V_{\text{CW}}(v_\chi \omega^j),
\end{equation}
which acts as a bias term and leads to the collapse of domain-walls. The finite-temperature corrections can also generate a bias term in the presence of explicit $\mathbb{Z}_3$ breaking interactions. When the scalar field $\chi$ is coupled to particles in the thermal bath, the effective potential receives temperature-dependent contributions that lift the degeneracy among the $\mathbb{Z}_3$ vacua.
The one-loop finite-temperature correction to the effective potential is given by~\cite{Dolan:1973qd,Quiros:1999jp}
\begin{equation}
V_T(\chi) = - \sum_i \frac{n_i T^4}{2\pi^2} \, J_F\!\left(\frac{m_i^2(\chi)}{T^2}\right),
\end{equation}
where $n_i$ denotes the degrees of freedom and $J_F$ is the fermionic thermal function. In the high-temperature limit, $m_i^2(\chi) \ll T^2$, this expression can be expanded to yield the leading contribution
\begin{equation}
V_T(\chi) \simeq \sum_i \frac{n_i}{48} \, m_i^2(\chi)\, T^2.
\end{equation}
\noindent Due to the field-dependent masses $m_i(\chi) = M_i + y_i \chi$, the thermal potential evaluated at these vacua becomes non-degenerate once explicit $\mathbb{Z}_3$ breaking is taken into account. This results in a temperature-dependent energy splitting between the minima,
\begin{equation}
\Delta V_T^{ij} = V_T(v_\chi \omega^i) - V_T(v_\chi \omega^j) \propto T^2,
\end{equation}
which acts as a thermal bias term. The finite-temperature contribution further enhances the dynamical lifting of the vacuum degeneracy during the early stages of cosmological evolution. Since the thermal correction scales approximately as $T^2$, the pressure difference across the walls becomes increasingly significant at high temperatures, thereby accelerating the collapse of the domain-wall network. Nevertheless, in the region of parameter space compatible with successful thermal leptogenesis, the zero-temperature Coleman-Weinberg contribution remains the leading correction to the scalar potential near the annihilation epoch. As a result, the annihilation process is closely linked to the radiative effects induced by the heavy fermion triplet sector, which also governs neutrino mass generation and CP asymmetry production.
Such a bias enhances the pressure difference across domain-walls and can significantly accelerate their collapse in the early Universe. The total bias between the $\mathbb{Z}_3$ vacua at finite temperature arises from both quantum and thermal effects, and can be expressed as
\begin{equation}
V_{\text{bias}}(T) = \Delta V_{\text{CW}} + \Delta V_T(T),
\label{total_bias}
\end{equation}
where $\Delta V_{\text{CW}}$ denotes the zero-temperature Coleman--Weinberg contribution, while $\Delta V_T(T)$ represents the finite-temperature correction. The Eqn.~\ref{total_bias} demonstrates that the vacuum degeneracy among the three $\mathbb{Z}_3$ minima is lifted dynamically through quantum effects induced by the heavy fermion triplets and the scalar singlet $\chi$. Unlike conventional approaches, where the bias term is introduced phenomenologically through higher-dimensional operators with arbitrarily chosen coefficients, the present framework generates the required bias radiatively from the Type-III seesaw sector itself. In the parameter space of interest, the zero-temperature contribution typically dominates at the time of domain-wall annihilation, i.e., $T \simeq T_{\text{ann}}$.
The evolution of domain-walls is governed by the competition between the tension force and the pressure difference induced by the bias. The tension force per unit area is given by
\begin{equation}
p_T \sim \frac{\sigma}{L},
\end{equation}
where the wall tension is $\sigma \sim \sqrt{\lambda}\, v_\chi^3$ and the characteristic wall width is $L \sim (\sqrt{\lambda}\, v_\chi)^{-1}$. 
As the Universe cools, domain-walls annihilate when the pressure due to the bias becomes comparable to the tension force, $i.e.$,
\begin{equation}
p_T \sim p_V.
\end{equation}
This condition determines the annihilation temperature $T_{\text{ann}}$. For the present model, it can be estimated as~\cite{Borah:2025bfa}
\begin{equation}
T_{\rm ann} =
\frac{
3.39 \times 10^{-2}
(C_{\rm ann}A)^{-1/2}
\left(\frac{g_*}{10}\right)^{-1/4}
\left(\frac{\sigma}{10^{9}\,\mathrm{GeV}^3}\right)^{-1/2}
\left(\frac{\Delta V_{\rm CW}}{10^{-12}\,\mathrm{GeV}^4}\right)^{1/2}
}{
\sqrt{
1-1.14\times10^{-3}
(C_{\rm ann}A)^{-1}
\left(\frac{g_*}{10}\right)^{-1/2}
\left(\frac{\sigma}{10^{9}\,\mathrm{GeV}^3}\right)^{-1}
\left(\frac{c_1}{10^{-12}\,\mathrm{GeV}^2}\right)
}
}
~\mathrm{GeV},
\label{annihilation_temp}
\end{equation}

\noindent where $C_{\text{ann}}$ and $A$ are model-dependent parameters encoding the dynamics of the domain-wall network. If the bias term is absent or sufficiently small, the domain-wall network eventually dominates the energy density of the Universe. The corresponding domination temperature is approximately given by~\cite{Saikawa:2017hiv, Borah:2025bfa}
\begin{equation}
T_{\rm dom} =
2.86 \times 10^{-6}
\left(\frac{g_*}{10}\right)^{-1/4}
\left(\frac{A}{0.8}\right)^{1/2}
\left(\frac{\sigma}{10^{9}\,\mathrm{GeV}^3}\right)^{1/2}
~\mathrm{GeV}.
\end{equation}

\noindent Therefore, in order to avoid the overclosure of the Universe, the annihilation of the domain-wall network must occur before the domination epoch, i.e.
\begin{equation}
T_{\rm ann} > T_{\rm dom}.
\end{equation}
This condition imposes a lower bound on the radiatively generated bias term. An additional lower bound on the bias term can be obtained by demanding that the domain-walls annihilate before the onset of Big Bang Nucleosynthesis (BBN),
\begin{equation}
T_{\rm ann} > T_{\rm BBN},
\end{equation}
so that the light-element abundances remain unaffected.

\noindent On the other hand, the bias term cannot be arbitrarily large. In the context of the present $\mathbb{Z}_3$ framework, an excessively large vacuum-energy splitting among the three vacua would prevent the simultaneous formation and percolation of the different $\mathbb{Z}_3$ domains in the early Universe. Consequently, the bias term is constrained from above as
\begin{equation}
V_{\rm bias} \lesssim 0.795\,V_0,
\end{equation}
where $V_0$ denotes the height of the potential barrier separating the different $\mathbb{Z}_3$ vacua. The annihilation of unstable domain-walls can produce a stochastic background of gravitational waves~\cite{Saikawa:2017hiv,Hiramatsu:2010yz,Hiramatsu:2013qaa}. Assuming that the $\mathbb{Z}_3$ domain-wall network annihilates instantaneously at $T=T_{\rm ann}$ during the radiation-dominated era, the present-day peak frequency $f_p$ and the corresponding peak energy-density spectrum $\Omega_p h^2$ of the resulting gravitational-wave signal can be approximately expressed as~\cite{Saikawa:2017hiv,Hiramatsu:2013qaa}



\begin{equation}
f_p \simeq 7.5 \times 10^{-9}\, \mathrm{Hz}
\left(\frac{T_{\rm gw}}{0.1\,\mathrm{GeV}}\right)
\left(\frac{g_*(T_{\rm gw})}{10}\right)^{1/6},
\end{equation}

\begin{equation}
\Omega_p h^2 \simeq 1 \times 10^{-10}
\left(\frac{0.1\,\mathrm{GeV}}{T_{\rm gw}}\right)^4
\left(\frac{\sigma}{10^{15}\,\mathrm{GeV}^3}\right)^2
\left(\frac{10}{g_*(T_{\rm gw})}\right)^{4/3}.
\end{equation}

\noindent Recent numerical studies \cite{Kitajima:2023cek,Notari:2025kqq} have shown that the peak gravitational-wave emission occurs at a temperature lower than the domain-wall annihilation temperature, namely ($T_{\rm gw} \simeq 0.3,T_{\rm ann}$). The stochastic gravitational-wave spectrum generated from the annihilation of the unstable $\mathbb{Z}_3$ domain-wall network can be described by a broken power-law spectrum~\cite{Caprini:2019egz,NANOGrav:2023hvm}. Following the standard scaling approximation for domain-wall evolution, the present-day gravitational-wave spectrum can be parametrized as~\cite{Hiramatsu:2013qaa,Saikawa:2017hiv}
\begin{equation}
\Omega_{\rm GW} h^2
=
\Omega_p h^2
\left[
\frac{(a+b)^c}
{b\left(\frac{f}{f_p}\right)^{-a/c}
+
a\left(\frac{f}{f_p}\right)^{b/c}}
\right]^c,
\end{equation}
where $f_p$ and $\Omega_p h^2$ denote the peak frequency and peak gravitational-wave energy density spectrum, respectively. Here, $a$, $b$, and $c$ are positive constants controlling the low- and high-frequency behavior of the spectrum. In the present analysis, we adopt the commonly used values $a=3, b, c=1,$~\cite{Hiramatsu:2013qaa,Borah:2025bfa} which reproduce the characteristic low-frequency causal behavior $\Omega_{\rm GW}\propto f^3$ and the high-frequency fall-off $\Omega_{\rm GW}\propto f^{-1}$ expected from domain-wall annihilation. Although the present framework is based on a $\mathbb{Z}_3$ symmetry rather than the simpler $\mathbb{Z}_2$ case, the gravitational-wave spectrum is expected to follow a similar broken power-law structure once the domain-wall network enters the scaling regime and annihilates during the radiation-dominated era. In particular, the peak position and amplitude are primarily controlled by the annihilation temperature and the radiatively generated bias term. Since the bias term originates from the Yukawa interactions between the scalar singlet and the heavy fermion triplets, the resulting gravitational-wave observables remain directly correlated with the Type-III seesaw and thermal leptogenesis parameters.
\begin{figure}
    \centering
    \includegraphics[width=0.7\linewidth]{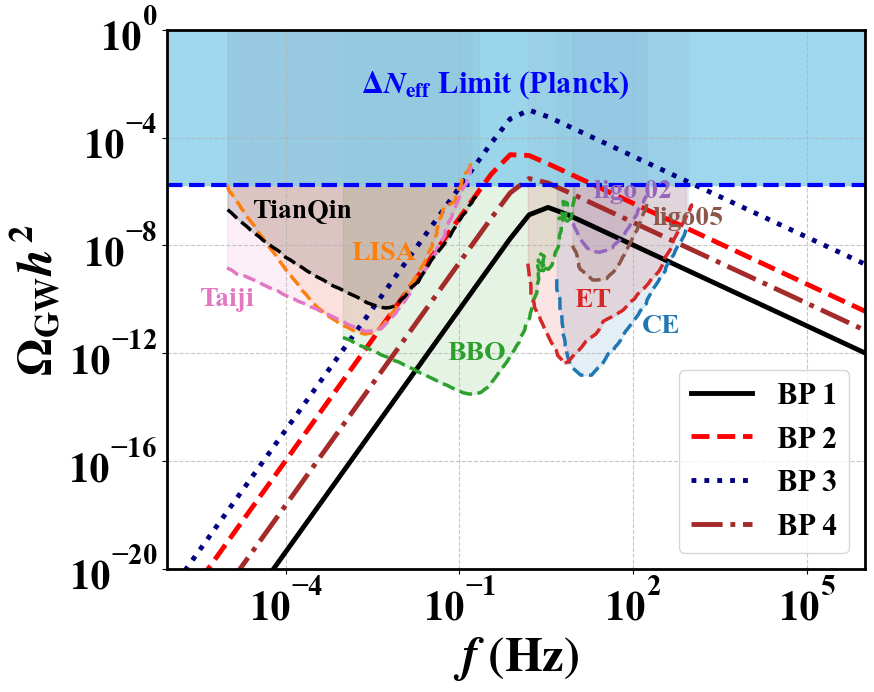}
    \caption{Gravitational wave peak amplitude as a function of the peak frequency for the selected benchmark points. The black, red, blue and brown curves represent BP1, BP2, BP3, and BP4 respectively. All benchmark points satisfy the current neutrino oscillation data and successfully reproduce the required $B-L$ asymmetry through thermal leptogenesis within the Type-III seesaw framework.}
    \label{gw_plot}
\end{figure}
Figure~\ref{gw_plot} shows the GW spectra for various model parameters. Depending on the scale of $\mathbb{Z}_3$-breaking, the seesaw scale, and the $\chi$-$\Sigma_i$ couplings, the peak amplitude of the spectrum can fall within the sensitivity reach of several present and future experiments. The dotted lines indicate the projected sensitivities of future GW detectors, including TianQin~\cite{TianQin:2015yph,TianQin:2020hid}, Taiji~\cite{Hu:2017mde, Ruan:2018tsw}, LISA~\cite{LISA:2017pwj}, BBO~\cite{Crowder:2005nr},ET~\cite{Punturo:2010zz,Hild:2010id,Sathyaprakash:2012jk}, LIGO~\cite{LIGOScientific:2014qfs,LIGOScientific:2016aoc,LIGOScientific:2016wof,LIGOScientific:2017vwq}, and CE~\cite{Reitze:2019iox}. The blue region in Fig.~\ref{gw_plot} indicates the $\Delta N_{eff}$ limit. Our results are largely compatible with the sensitivity bands of space-based and ground-based GW experiments. Since reproducing the observed neutrino masses and generating sufficient $B$-$L$ asymmetry requires a heavy triplet, we work with $M_1 \sim 10^{9}$ GeV. This requirement has important implications for the gravitational-wave phenomenology of the model. Since the same heavy fermions generate the radiative $\mathbb{Z}_3$-breaking bias term, the leptogenesis constraint naturally pushes the domain-wall annihilation temperature toward higher scales, thereby shifting the gravitational-wave signal toward higher frequencies.

\noindent To ensure consistency with standard cosmological evolution, the domain-walls must annihilate before they dominate the energy density of the Universe, which imposes a lower bound on the annihilation temperature. For the selected benchmark point BP1, the corresponding annihilation temperature obtained from Eqn.~\ref{annihilation_temp} is found to 
be $T_{\rm ann} = 8.74 \times10^{7}\,\mathrm{GeV}$, ensuring that the domain-walls annihilate before dominating the energy density of the Universe. The resulting gravitational-wave spectrum for the benchmark point is shown in Fig.~\ref{gw_plot}. Owing to the large $\mathbb{Z}_3$ symmetry-breaking scale and the high Type-III seesaw scale associated with successful leptogenesis, the peak of the gravitational-wave spectrum is shifted towards the high-frequency regime. For BP1, the peak frequency is obtained as $f_{\rm peak} = 2.92\,\mathrm{Hz}$ with a corresponding peak amplitude of $\Omega_{\rm GW} h^2 = 2.63 \times 10^{-7}$. The predicted spectrum fall within the sensitivity reach of future space and ground based gravitational-wave experiments such as BBO, ET, LIGO and CE. For the benchmark point BP2, the annihilation temperature obtained from Eqn.~\ref{annihilation_temp} is found to be 
$T_{\rm ann} = 3.04 \times 10^{7}\,\mathrm{GeV}$. The corresponding gravitational-wave spectrum is characterized by a peak frequency of 
$f_{\rm peak} = 1.01 \,\mathrm{Hz}$, 
with a peak amplitude of 
$\Omega_{\rm GW} h^2 = 2.69 \times 10^{-5}$. 
The predicted signal lies within the projected sensitivity range of Taiji, TianQin, LISA, BBO, LIGO, ET, and CE. For BP3, the annihilation temperature calculated from Eqn.~\ref{annihilation_temp} is 
$T_{\rm ann} = 4.31 \times 10^{7}\,\mathrm{GeV}$. The resulting gravitational-wave signal exhibits a peak frequency of 
$f_{\rm peak} = 1.43\,\mathrm{Hz}$ 
and a peak intensity of 
$\Omega_{\rm GW} h^2 = 1.05 \times 10^{-3}$. 
The corresponding spectrum is expected to be probed by future gravitational-wave experiments, particularly Taiji, LISA, TianQion, and BBO. For BP4, the annihilation temperature calculated from Eqn.~\ref{annihilation_temp} is 
$T_{\rm ann} = 5.23 \times 10^{7}\,\mathrm{GeV}$. The resulting gravitational-wave signal exhibits a peak frequency of 
$f_{\rm peak} = 1.74\,\mathrm{Hz}$ 
and a peak intensity of 
$\Omega_{\rm GW} h^2 = 3.06 \times 10^{-6}$. 
The corresponding spectrum is expected to be probed by future gravitational-wave experiments, particularly BBO, LIGO, ET and CE. 

\noindent Fig.~\ref{gw_plot} illustrates a clear correlation can be observed between the symmetry-breaking scale ($v_\chi$), the mass of the heavy fermion triplets, and the resulting gravitational-wave signal. In particular, benchmark points associated with larger values of $v_\chi$ generally exhibit higher peak frequencies and larger peak amplitudes. This behavior originates from the fact that the domain-wall tension scales approximately as $\sigma \propto v_\chi^3$, while the radiatively generated bias term also increases with the symmetry-breaking scale through the field-dependent fermion masses. Consequently, the domain-walls annihilate at higher temperatures, shifting the gravitational-wave peak towards higher frequencies.

\noindent The benchmark points presented in Table~\ref{tab:BPs} simultaneously satisfy neutrino oscillation constraints, successful thermal leptogenesis, and the cosmological requirements associated with unstable domain-walls. In particular, all benchmark points reproduce the observed neutrino mixing parameters within the allowed experimental ranges while generating sufficient baryon asymmetry through the decay of the lightest fermion triplet. At the same time, the corresponding radiatively generated bias terms ensure sufficiently early annihilation of the domain-wall network, leading to potentially observable gravitational-wave signatures. Since the fermion triplet masses required for successful Type-III leptogenesis lie around $10^{9}~\mathrm{GeV}$, they remain far beyond the direct reach of present and future collider experiments. Nevertheless, in the present framework, the same heavy fermion sector controls the radiative lifting of the vacuum degeneracy and hence determines the domain-wall annihilation dynamics. Consequently, the resulting gravitational-wave spectrum provides an indirect probe of the high-scale leptogenesis and Type-III seesaw framework through its correlation with the annihilation temperature and the radiatively generated bias term.
\section{Conclusions}\label{section6}
\noindent In this work, we have investigated a $\mathbb{Z}_3$-symmetric extension of the Type-III seesaw framework containing three hyperchargeless $SU(2)_L$ fermion triplets and a complex scalar singlet $\chi$. After the spontaneous breaking of the discrete $\mathbb{Z}_3$ symmetry, the scalar singlet develops three degenerate vacua, leading to the formation of domain-walls in the early Universe. The neutrino masses are generated through the Type-III seesaw mechanism, while the decay of the lightest fermion triplet produces the observed baryon asymmetry through thermal leptogenesis. A central feature of the present framework is that the explicit $\mathbb{Z}_3$-breaking bias responsible for domain-wall annihilation is not introduced phenomenologically. Instead, it is generated dynamically through radiative corrections induced by the Yukawa interactions between the scalar singlet and the heavy fermion triplets. Consequently, the same sector responsible for neutrino mass generation and leptogenesis also governs the instability and annihilation of the domain-wall network. This establishes a direct connection between neutrino physics, baryogenesis, domain-wall dynamics, and gravitational-wave phenomenology.\\
We performed a numerical analysis consistent with the latest neutrino oscillation data and identified viable regions of parameter space compatible with the observed neutrino masses and leptonic mixing parameters. The framework successfully generates the observed baryon asymmetry through thermal leptogenesis for fermion triplet masses around $10^{9}~\mathrm{GeV}$. In the Type-III seesaw scenario, electroweak gauge interactions of the fermion triplets enhance the washout processes, thereby favoring a relatively high leptogenesis scale. Since the same heavy fermion sector also generates the radiative $\mathbb{Z}_3$-breaking bias term, the domain-wall annihilation temperature is naturally shifted toward high scales.\\
The annihilation of unstable domain-walls produces a stochastic gravitational-wave background whose peak frequency lies in the high-frequency regime. We showed that the predicted gravitational-wave spectra corresponding to the selected benchmark points can fall within the projected sensitivity reach of future ground-based and space based gravitational-wave experiments such as TianQin~\cite{TianQin:2015yph,TianQin:2020hid}, Taiji~\cite{Hu:2017mde, Ruan:2018tsw}, LISA~\cite{LISA:2017pwj}, BBO~\cite{Crowder:2005nr},ET~\cite{Punturo:2010zz,Hild:2010id,Sathyaprakash:2012jk}, LIGO~\cite{LIGOScientific:2014qfs,LIGOScientific:2016aoc,LIGOScientific:2016wof,LIGOScientific:2017vwq}, and CE~\cite{Reitze:2019iox}. Therefore, although the heavy fermion triplets themselves remain far beyond the direct reach of collider experiments, the resulting gravitational-wave signal provides an indirect cosmological probe of the high-scale Type-III seesaw and leptogenesis framework. Unlike conventional approaches where the domain-wall bias term is introduced ad hoc, the present framework dynamically relates the origin of the bias term to the same interactions responsible for neutrino mass generation. As a result, the gravitational-wave observables become directly correlated with the scales governing leptogenesis and the Type-III seesaw mechanism, leading to a more predictive and constrained scenario. The present work therefore demonstrates that future gravitational-wave observations can provide an important window into otherwise inaccessible high-scale neutrino mass generation and baryogenesis physics.\\
The present work provides one of the first ultraviolet-complete realizations connecting neutrino phenomenology, thermal leptogenesis, unstable domain-wall dynamics, and gravitational-wave production within a Type-III seesaw framework. In contrast to phenomenological approaches where the gravitational-wave sector is often treated independently, the present framework simultaneously accommodates the observed neutrino oscillation data through a detailed parameter-space analysis while dynamically generating the domain-wall bias term through the same fermion sector responsible for neutrino mass generation. Consequently, the predicted gravitational-wave observables are directly correlated with the underlying neutrino and leptogenesis parameters, leading to a more constrained and predictive cosmological scenario.
\section*{Acknowledgments} 
\noindent Priya, B. C. Chauhan and Surender Verma would like to acknowledge the IUCAA for providing the HPC facility to carry out this work. L. S. acknowledges the financial support provided by the Council of Scientific and Industrial Research (CSIR) vide letter No. 09/1196(18553)/2024-EMR-I. L. S. would also like to thank the organizers of ``Workshop on High Energy Physics Phenomenology 2025 (WHEPP-2025)" at IIT Hyderabad, where part of this work was discussed. 

\appendix
\section{Yukawa coupling matrix $y_d$}\label{appendix1}
\begin{equation}
    y_d = \begin{pmatrix}
        0.0012-0.0033 & 0.0017-0.0037 & 0.0017-0.0037 \\
        0.0017-0.0037 &0.0019 - 0.0073 &0.0017-0.0037\\
        0.0017-0.0037 & 0.0017-0.0037 & 0.0017-0.0036
    \end{pmatrix}
    \label{yd}
\end{equation}

\noindent The range of Yukawa couplings $y_d$ used in the present analysis is given in Eqn. \ref{yd}. These couplings play an important role in determining the efficiency of thermal leptogenesis through the decay of the heavy fermion triplets and simultaneously contribute to the radiatively generated $\mathbb{Z}_3$-breaking bias term responsible for domain-wall annihilation. The selected parameter ranges are consistent with the neutrino oscillation constraints and lead to successful generation of the observed $B-L$ asymmetry. Since the same Yukawa interactions also enter the Coleman--Weinberg effective potential, they directly influence the annihilation temperature of the domain-wall network and consequently the resulting gravitational-wave spectrum.

\section{Gauge interactions and inverse decay rate}\label{appendix2}

\begin{figure}
    \centering
    \includegraphics[width=0.49\linewidth]{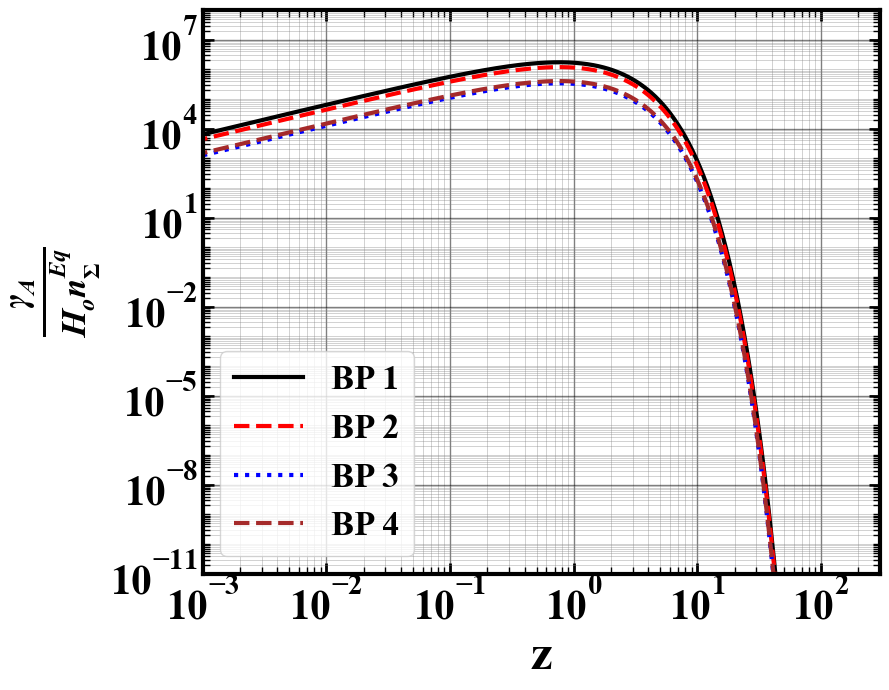}
    \includegraphics[width=0.49\linewidth]{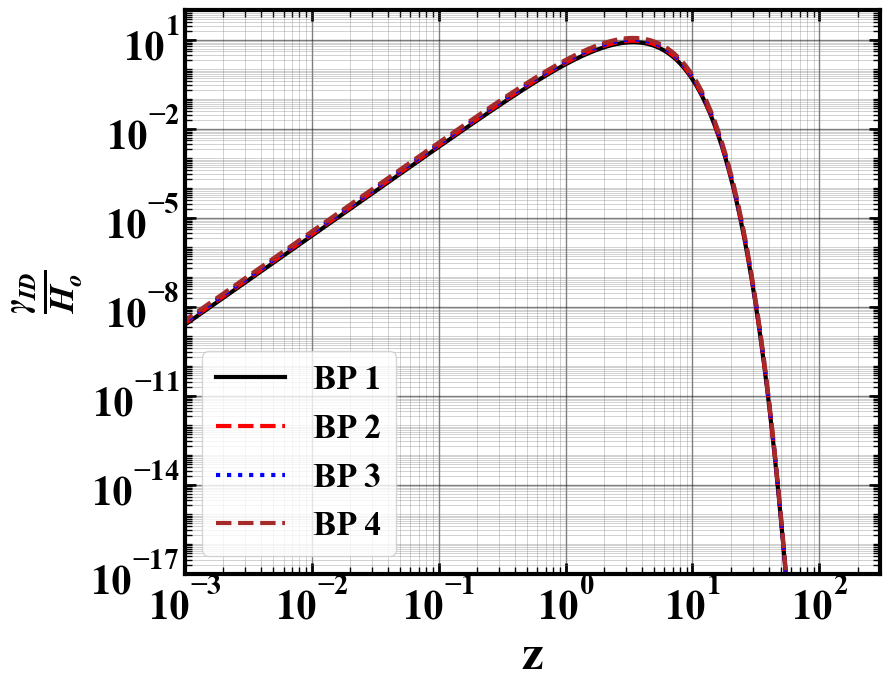}
    \caption{The gauge interactions (left panel) and inverse decay (right panel) rate as function of $z=m_{1}/T$ for BPs given in Table \ref{tab:BPs}.}
    \label{decay_processes}
\end{figure}
\noindent The left panel of Fig. \ref{decay_processes} shows the evolution of the gauge scattering reaction density, $\frac{\gamma_A}{H_0\, n_{\Sigma}^{\rm Eq}}$, as a function of $z=m_{1}/T$. The reaction rate remains large around $z \lesssim 1$, indicating that gauge interactions efficiently keep the fermion triplets in thermal equilibrium at early times. As $z$ increases further, the reaction density becomes Boltzmann suppressed and rapidly decreases. The nearly identical behavior for all benchmark points shows that gauge scatterings are mainly controlled by the $SU(2)_L$ gauge interactions and are weakly dependent on the benchmark parameters. On the other hand, the right panel of Fig ~4 depicts the decay reaction density, $\frac{\gamma_{ID}}{H_0}$, for the four benchmark points. The decay processes become significant near $z \sim 1$, corresponding to the epoch when the heavy triplets become non-relativistic. At larger $z$, the decay rate decreases sharply due to the suppression in the triplet abundance. These out-of-equilibrium decays are responsible for generating the lepton asymmetry required for successful leptogenesis in the Type-III seesaw framework.

\FloatBarrier

\bibliographystyle{JHEP}
\bibliography{ref.bib} 

@article{Kitajima:2023cek,
    author = "Kitajima, Naoya and Lee, Junseok and Murai, Kai and Takahashi, Fuminobu and Yin, Wen",
    title = "{Gravitational waves from domain wall collapse, and application to nanohertz signals with QCD-coupled axions}",
    eprint = "2306.17146",
    archivePrefix = "arXiv",
    primaryClass = "hep-ph",
    reportNumber = "TU-1198",
    doi = "10.1016/j.physletb.2024.138586",
    journal = "Phys. Lett. B",
    volume = "851",
    pages = "138586",
    year = "2024"
}

@article{Notari:2025kqq,
    author = "Notari, Alessio and Rompineve, Fabrizio and Torrenti, Francisco",
    title = "{The spectrum of gravitational waves from annihilating domain walls}",
    eprint = "2504.03636",
    archivePrefix = "arXiv",
    primaryClass = "astro-ph.CO",
    doi = "10.1088/1475-7516/2025/07/049",
    journal = "JCAP",
    volume = "07",
    pages = "049",
    year = "2025"
}

@article{Coleman:1973jx,
    author = "Coleman, Sidney R. and Weinberg, Erick J.",
    title = "{Radiative Corrections as the Origin of Spontaneous Symmetry Breaking}",
    doi = "10.1103/PhysRevD.7.1888",
    journal = "Phys. Rev. D",
    volume = "7",
    pages = "1888--1910",
    year = "1973"
}

@inproceedings{Quiros:1999jp,
    author = "Quiros, Mariano",
    title = "{Finite temperature field theory and phase transitions}",
    booktitle = "{ICTP Summer School in High-Energy Physics and Cosmology}",
    eprint = "hep-ph/9901312",
    archivePrefix = "arXiv",
    reportNumber = "IEM-FT-187-99",
    pages = "187--259",
    month = "1",
    year = "1999"
}

@article{Dolan:1973qd,
    author = "Dolan, L. and Jackiw, R.",
    title = "{Symmetry Behavior at Finite Temperature}",
    reportNumber = "MIT-CTP-406",
    doi = "10.1103/PhysRevD.9.3320",
    journal = "Phys. Rev. D",
    volume = "9",
    pages = "3320--3341",
    year = "1974"
}

@article{Caprini:2019egz,
    author = "Caprini, Chiara and others",
    title = "{Detecting gravitational waves from cosmological phase transitions with LISA: an update}",
    eprint = "1910.13125",
    archivePrefix = "arXiv",
    primaryClass = "astro-ph.CO",
    reportNumber = "DESY-19-159, IPPP/19/27, HIP-2019-14/TH, MITP/19-066, IFT-UAM/CSIC-19-139",
    doi = "10.1088/1475-7516/2020/03/024",
    journal = "JCAP",
    volume = "03",
    pages = "024",
    year = "2020"
}

@article{NANOGrav:2023hvm,
    author = "Afzal, Adeela and others",
    collaboration = "NANOGrav",
    title = "{The NANOGrav 15 yr Data Set: Search for Signals from New Physics}",
    eprint = "2306.16219",
    archivePrefix = "arXiv",
    primaryClass = "astro-ph.HE",
    reportNumber = "FERMILAB-PUB-23-589-T",
    doi = "10.3847/2041-8213/acdc91",
    journal = "Astrophys. J. Lett.",
    volume = "951",
    number = "1",
    pages = "L11",
    year = "2023",
    note = "[Erratum: Astrophys.J.Lett. 971, L27 (2024), Erratum: Astrophys.J. 971, L27 (2024)]"
}

@article{Kibble:1976sj,
    author = "Kibble, T. W. B.",
    title = "{Topology of Cosmic Domains and Strings}",
    reportNumber = "ICTP/75/5",
    doi = "10.1088/0305-4470/9/8/029",
    journal = "J. Phys. A",
    volume = "9",
    pages = "1387--1398",
    year = "1976"
}

@article{Zeldovich:1974uw,
    author = "Zeldovich, Ya. B. and Kobzarev, I. Yu. and Okun, L. B.",
    title = "{Cosmological Consequences of the Spontaneous Breakdown of Discrete Symmetry}",
    reportNumber = "SLAC-TRANS-0165, IPM-MOSCOW-15",
    journal = "Zh. Eksp. Teor. Fiz.",
    volume = "67",
    pages = "3--11",
    year = "1974"
}

@article{Barreto:2026oyf,
    author = "Barreto, Gon{\c{c}}alo and de Medeiros Varzielas, Ivo and Zhou, Ye-Ling",
    title = "{Domain walls from \(\sum(36 \times 3)\), \(\Delta(54)\) and \(\Delta(27)\) potentials}",
    doi = "10.1103/7qx6-568y",
    journal = "Phys. Rev. D",
    volume = "113",
    number = "11",
    pages = "115019",
    year = "2026"
}

@article{King:2024lki,
    author = "King, Stephen F. and Wang, Xin and Zhou, Ye-Ling",
    title = "{Modular domain walls and gravitational waves}",
    eprint = "2411.04900",
    archivePrefix = "arXiv",
    primaryClass = "hep-ph",
    doi = "10.1088/1475-7516/2025/07/011",
    journal = "JCAP",
    volume = "07",
    pages = "011",
    year = "2025"
}

@article{Winckler:2025hbc,
    author = "Winckler, Clara and Avelino, Pedro P. and Sousa, Lara",
    title = "{Biased domain walls and the origin of early massive structures}",
    eprint = "2507.22812",
    archivePrefix = "arXiv",
    primaryClass = "astro-ph.CO",
    doi = "10.1103/kp7r-2b3l",
    journal = "Phys. Rev. D",
    volume = "112",
    number = "12",
    pages = "123506",
    year = "2025"
}

@article{Zeng:2025zjp,
    author = "Zeng, Qing-Quan and He, Xi and Yu, Zhao-Huan and Zheng, Jiaming",
    title = "{Collapsing domain walls with Z2-violating coupling to thermalized fermions and their impact on gravitational wave detections}",
    eprint = "2501.10059",
    archivePrefix = "arXiv",
    primaryClass = "hep-ph",
    doi = "10.1103/cdsj-bmvx",
    journal = "Phys. Rev. D",
    volume = "111",
    number = "11",
    pages = "115017",
    year = "2025"
}

@article{AristizabalSierra:2010mv,
    author = "Aristizabal Sierra, D. and Kamenik, Jernej F. and Nemevsek, Miha",
    title = "{Implications of Flavor Dynamics for Fermion Triplet Leptogenesis}",
    eprint = "1007.1907",
    archivePrefix = "arXiv",
    primaryClass = "hep-ph",
    doi = "10.1007/JHEP10(2010)036",
    journal = "JHEP",
    volume = "10",
    pages = "036",
    year = "2010"
}

@article{Priya:2026ehe,
    author = "Priya and Chauhan, B. C. and Kumar, Deepak and Nomura, Takaaki",
    title = "{Predictions of Modular Symmetry Fixed Points on Neutrino Masses, Mixing, and Leptogenesis}",
    eprint = "2604.04585",
    archivePrefix = "arXiv",
    primaryClass = "hep-ph",
    month = "4",
    year = "2026"
}

@article{Fukugita:1986hr,
    author = "Fukugita, M. and Yanagida, T.",
    title = "{Baryogenesis Without Grand Unification}",
    reportNumber = "RIFP-641",
    doi = "10.1016/0370-2693(86)91126-3",
    journal = "Phys. Lett. B",
    volume = "174",
    pages = "45--47",
    year = "1986"
}

@article{Covi:1996wh,
    author = "Covi, Laura and Roulet, Esteban and Vissani, Francesco",
    title = "{CP violating decays in leptogenesis scenarios}",
    eprint = "hep-ph/9605319",
    archivePrefix = "arXiv",
    reportNumber = "SISSA-66-96-EP, IC-96-73",
    doi = "10.1016/0370-2693(96)00817-9",
    journal = "Phys. Lett. B",
    volume = "384",
    pages = "169--174",
    year = "1996"
}

@article{Buchmuller:2005eh,
    author = "Buchmuller, W. and Peccei, R. D. and Yanagida, T.",
    title = "{Leptogenesis as the origin of matter}",
    eprint = "hep-ph/0502169",
    archivePrefix = "arXiv",
    reportNumber = "DESY-05-031",
    doi = "10.1146/annurev.nucl.55.090704.151558",
    journal = "Ann. Rev. Nucl. Part. Sci.",
    volume = "55",
    pages = "311--355",
    year = "2005"
}

@article{Sakharov:1967dj,
    author = "Sakharov, A. D.",
    title = "{Violation of CP Invariance, C asymmetry, and baryon asymmetry of the universe}",
    doi = "10.1070/PU1991v034n05ABEH002497",
    journal = "Pisma Zh. Eksp. Teor. Fiz.",
    volume = "5",
    pages = "32--35",
    year = "1967"
}

@article{Kuzmin:1985mm,
    author = "Kuzmin, V. A. and Rubakov, V. A. and Shaposhnikov, M. E.",
    title = "{On the Anomalous Electroweak Baryon Number Nonconservation in the Early Universe}",
    reportNumber = "IC/85/8",
    doi = "10.1016/0370-2693(85)91028-7",
    journal = "Phys. Lett. B",
    volume = "155",
    pages = "36",
    year = "1985"
}

@article{Buchmuller:2004nz,
    author = "Buchmuller, W. and Di Bari, P. and Plumacher, M.",
    title = "{Leptogenesis for pedestrians}",
    eprint = "hep-ph/0401240",
    archivePrefix = "arXiv",
    reportNumber = "DESY-03-100, UAB-FT-551, CERN-TH-2003-199",
    doi = "10.1016/j.aop.2004.02.003",
    journal = "Annals Phys.",
    volume = "315",
    pages = "305--351",
    year = "2005"
}

@article{Okada:2018xdh,
    author = "Okada, Nobuchika and Seto, Osamu",
    title = "{Probing the seesaw scale with gravitational waves}",
    eprint = "1807.00336",
    archivePrefix = "arXiv",
    primaryClass = "hep-ph",
    reportNumber = "EPHOU-18-007",
    doi = "10.1103/PhysRevD.98.063532",
    journal = "Phys. Rev. D",
    volume = "98",
    number = "6",
    pages = "063532",
    year = "2018"
}

@article{Hasegawa:2019amx,
    author = "Hasegawa, Taiki and Okada, Nobuchika and Seto, Osamu",
    title = "{Gravitational waves from the minimal gauged $U(1)_{B-L}$ model}",
    eprint = "1904.03020",
    archivePrefix = "arXiv",
    primaryClass = "hep-ph",
    reportNumber = "EPHOU-19-004",
    doi = "10.1103/PhysRevD.99.095039",
    journal = "Phys. Rev. D",
    volume = "99",
    number = "9",
    pages = "095039",
    year = "2019"
}

@article{Borah:2022cdx,
    author = "Borah, Debasish and Dasgupta, Arnab and Saha, Indrajit",
    title = "{Leptogenesis and dark matter through relativistic bubble walls with observable gravitational waves}",
    eprint = "2207.14226",
    archivePrefix = "arXiv",
    primaryClass = "hep-ph",
    doi = "10.1007/JHEP11(2022)136",
    journal = "JHEP",
    volume = "11",
    pages = "136",
    year = "2022"
}

@article{Borah:2023saq,
    author = "Borah, Debasish and Dasgupta, Arnab and Saha, Indrajit",
    title = "{LIGO-Virgo constraints on dark matter and leptogenesis triggered by a first order phase transition at high scale}",
    eprint = "2304.08888",
    archivePrefix = "arXiv",
    primaryClass = "hep-ph",
    doi = "10.1103/PhysRevD.109.095034",
    journal = "Phys. Rev. D",
    volume = "109",
    number = "9",
    pages = "095034",
    year = "2024"
}

@article{Barman:2023fad,
    author = "Barman, Basabendu and Borah, Debasish and Jyoti Das, Suruj and Saha, Indrajit",
    title = "{Scale of Dirac leptogenesis and left-right symmetry in the light of recent PTA results}",
    eprint = "2307.00656",
    archivePrefix = "arXiv",
    primaryClass = "hep-ph",
    doi = "10.1088/1475-7516/2023/10/053",
    journal = "JCAP",
    volume = "10",
    pages = "053",
    year = "2023"
}

@article{Borah:2022vsu,
    author = "Borah, Debasish and Jyoti Das, Suruj and Roshan, Rishav",
    title = "{Probing high scale seesaw and PBH generated dark matter via gravitational waves with multiple tilts}",
    eprint = "2208.04965",
    archivePrefix = "arXiv",
    primaryClass = "hep-ph",
    doi = "10.1016/j.nuclphysb.2024.116528",
    journal = "Nucl. Phys. B",
    volume = "1002",
    pages = "116528",
    year = "2024"
}

@article{Dasgupta:2022isg,
    author = "Dasgupta, Arnab and Dev, P. S. Bhupal and Ghoshal, Anish and Mazumdar, Anupam",
    title = "{Gravitational wave pathway to testable leptogenesis}",
    eprint = "2206.07032",
    archivePrefix = "arXiv",
    primaryClass = "hep-ph",
    doi = "10.1103/PhysRevD.106.075027",
    journal = "Phys. Rev. D",
    volume = "106",
    number = "7",
    pages = "075027",
    year = "2022"
}

@article{Hambye:2012fh,
    author = "Hambye, Thomas",
    title = "{Leptogenesis: beyond the minimal type I seesaw scenario}",
    eprint = "1212.2888",
    archivePrefix = "arXiv",
    primaryClass = "hep-ph",
    reportNumber = "ULB-TH-12-13",
    doi = "10.1088/1367-2630/14/12/125014",
    journal = "New J. Phys.",
    volume = "14",
    pages = "125014",
    year = "2012"
}

@article{Albright:2003xb,
    author = "Albright, Carl H. and Barr, S. M.",
    title = "{Leptogenesis in the type III seesaw mechanism}",
    eprint = "hep-ph/0312224",
    archivePrefix = "arXiv",
    reportNumber = "FERMILAB-PUB-03-405-T, BA-03-21",
    doi = "10.1103/PhysRevD.69.073010",
    journal = "Phys. Rev. D",
    volume = "69",
    pages = "073010",
    year = "2004"
}

@article{Samanta:2020cdk,
    author = "Samanta, Rome and Datta, Satyabrata",
    title = "{Gravitational wave complementarity and impact of NANOGrav data on gravitational leptogenesis}",
    eprint = "2009.13452",
    archivePrefix = "arXiv",
    primaryClass = "hep-ph",
    doi = "10.1007/JHEP05(2021)211",
    journal = "JHEP",
    volume = "05",
    pages = "211",
    year = "2021"
}

@article{Barman:2022yos,
    author = "Barman, Basabendu and Borah, Debasish and Dasgupta, Arnab and Ghoshal, Anish",
    title = "{Probing high scale Dirac leptogenesis via gravitational waves from domain walls}",
    eprint = "2205.03422",
    archivePrefix = "arXiv",
    primaryClass = "hep-ph",
    doi = "10.1103/PhysRevD.106.015007",
    journal = "Phys. Rev. D",
    volume = "106",
    number = "1",
    pages = "015007",
    year = "2022"
}

@article{Huang:2022vkf,
    author = "Huang, Peisi and Xie, Ke-Pan",
    title = "{Leptogenesis triggered by a first-order phase transition}",
    eprint = "2206.04691",
    archivePrefix = "arXiv",
    primaryClass = "hep-ph",
    doi = "10.1007/JHEP09(2022)052",
    journal = "JHEP",
    volume = "09",
    pages = "052",
    year = "2022"
}

@article{Dror:2019syi,
    author = "Dror, Jeff A. and Hiramatsu, Takashi and Kohri, Kazunori and Murayama, Hitoshi and White, Graham",
    title = "{Testing the Seesaw Mechanism and Leptogenesis with Gravitational Waves}",
    eprint = "1908.03227",
    archivePrefix = "arXiv",
    primaryClass = "hep-ph",
    reportNumber = "IPMU19-0108, DESY-19-138, DESY 19-138, KEK-TH-2147, KEK-Cosmo-241",
    doi = "10.1103/PhysRevLett.124.041804",
    journal = "Phys. Rev. Lett.",
    volume = "124",
    number = "4",
    pages = "041804",
    year = "2020"
}

@article{Blasi:2020wpy,
    author = "Blasi, Simone and Brdar, Vedran and Schmitz, Kai",
    title = "{Fingerprint of low-scale leptogenesis in the primordial gravitational-wave spectrum}",
    eprint = "2004.02889",
    archivePrefix = "arXiv",
    primaryClass = "hep-ph",
    reportNumber = "CERN-TH-2020-055",
    doi = "10.1103/PhysRevResearch.2.043321",
    journal = "Phys. Rev. Res.",
    volume = "2",
    number = "4",
    pages = "043321",
    year = "2020"
}

@article{Fornal:2020esl,
    author = "Fornal, Bartosz and Shams Es Haghi, Barmak",
    title = "{Baryon and Lepton Number Violation from Gravitational Waves}",
    eprint = "2008.05111",
    archivePrefix = "arXiv",
    primaryClass = "hep-ph",
    doi = "10.1103/PhysRevD.102.115037",
    journal = "Phys. Rev. D",
    volume = "102",
    number = "11",
    pages = "115037",
    year = "2020"
}

@article{Borah:2025bfa,
    author = "Borah, Debasish and Saha, Indrajit",
    title = "{Gravitational waves from seesaw assisted collapsing domain walls}",
    eprint = "2512.22339",
    archivePrefix = "arXiv",
    primaryClass = "hep-ph",
    month = "12",
    year = "2025"
}

@article{Bhandari:2026ujy,
    author = "Bhandari, Dipendu and Borah, Debasish and Saha, Indrajit",
    title = "{Imprint of matter-antimatter asymmetry on collapsing domain walls}",
    eprint = "2604.02421",
    archivePrefix = "arXiv",
    primaryClass = "hep-ph",
    month = "4",
    year = "2026"
}

@article{Palavric:2024gvu,
    author = "Palavri{\'c}, Ajdin",
    title = "{Discrete leptonic flavor symmetries: UV mediators and phenomenology}",
    eprint = "2408.16044",
    archivePrefix = "arXiv",
    primaryClass = "hep-ph",
    doi = "10.1103/PhysRevD.110.115025",
    journal = "Phys. Rev. D",
    volume = "110",
    number = "11",
    pages = "115025",
    year = "2024"
}

@article{Blasi:2024vew,
    author = "Blasi, Simone and Calibbi, Lorenzo and Mariotti, Alberto and Turbang, Kevin",
    title = "{Gravitational waves from cosmic strings in Froggatt-Nielsen flavour models}",
    eprint = "2410.08668",
    archivePrefix = "arXiv",
    primaryClass = "hep-ph",
    reportNumber = "DESY-24-147, DESY--24--147",
    doi = "10.1007/JHEP05(2025)019",
    journal = "JHEP",
    volume = "05",
    pages = "019",
    year = "2025"
}

@article{Jana:2025vyb,
    author = "Jana, Sudip and Manna, Sudip and K, Vishnu P.",
    title = "{Gravitational Wave Signature and the Nature of Neutrino Masses: Majorana, Dirac, or Pseudo-Dirac?}",
    eprint = "2509.10456",
    archivePrefix = "arXiv",
    primaryClass = "hep-ph",
    reportNumber = "HRI-RECAPP-2025-09, MS-TP-25-19",
    doi = "10.1016/j.physletb.2026.140476",
    journal = "Phys. Lett. B",
    volume = "877",
    pages = "140476",
    year = "2026"
}

@article{Fu:2024jhu,
    author = "Fu, Bowen and King, Stephen F. and Marsili, Luca and Pascoli, Silvia and Turner, Jessica and Zhou, Ye-Ling",
    title = "{Non-Abelian domain walls and gravitational waves}",
    eprint = "2409.16359",
    archivePrefix = "arXiv",
    primaryClass = "hep-ph",
    reportNumber = "IPPP/24/62",
    doi = "10.1007/JHEP04(2025)142",
    journal = "JHEP",
    volume = "04",
    pages = "142",
    year = "2025"
}

@article{Wu:2022tpe,
    author = "Wu, Yongcheng and Xie, Ke-Pan and Zhou, Ye-Ling",
    title = "{Classification of Abelian domain walls}",
    eprint = "2205.11529",
    archivePrefix = "arXiv",
    primaryClass = "hep-ph",
    doi = "10.1103/PhysRevD.106.075019",
    journal = "Phys. Rev. D",
    volume = "106",
    number = "7",
    pages = "075019",
    year = "2022"
}

@article{Wei:2022poh,
    author = "Wei, Dongdong and Jiang, Yun",
    title = "{Domain wall networks from first-order phase transitions and gravitational waves}",
    eprint = "2208.07186",
    archivePrefix = "arXiv",
    primaryClass = "hep-ph",
    doi = "10.1103/PhysRevD.110.123505",
    journal = "Phys. Rev. D",
    volume = "110",
    number = "12",
    pages = "123505",
    year = "2024"
}

@article{Wu:2022stu,
    author = "Wu, Yongcheng and Xie, Ke-Pan and Zhou, Ye-Ling",
    title = "{Collapsing domain walls beyond Z2}",
    eprint = "2204.04374",
    archivePrefix = "arXiv",
    primaryClass = "hep-ph",
    doi = "10.1103/PhysRevD.105.095013",
    journal = "Phys. Rev. D",
    volume = "105",
    number = "9",
    pages = "095013",
    year = "2022"
}

@article{Barreto:2026igt,
    author = "Barreto, Gon{\c{c}}alo and de Medeiros Varzielas, Ivo and Zhou, Ye-Ling",
    title = "{Domain Walls from $\Sigma(36 \times 3)$, $\Delta(54)$ and $\Delta(27)$ potentials}",
    eprint = "2603.04496",
    archivePrefix = "arXiv",
    primaryClass = "hep-ph",
    month = "3",
    year = "2026"
}

@article{Li:2025gld,
    author = "Li, Yuan-Jie and Liu, Jing and Guo, Zong-Kuan",
    title = "{Dynamics of ZN domain walls with bias directions}",
    eprint = "2502.13644",
    archivePrefix = "arXiv",
    primaryClass = "astro-ph.CO",
    doi = "10.1103/rnpp-7wh2",
    journal = "Phys. Rev. D",
    volume = "112",
    number = "10",
    pages = "103510",
    year = "2025"
}

@article{Davidson:2002qv,
    author = "Davidson, Sacha and Ibarra, Alejandro",
    title = "{A Lower bound on the right-handed neutrino mass from leptogenesis}",
    eprint = "hep-ph/0202239",
    archivePrefix = "arXiv",
    reportNumber = "OUTP-02-10P, IPPP-02-16, DCPT-02-32",
    doi = "10.1016/S0370-2693(02)01735-5",
    journal = "Phys. Lett. B",
    volume = "535",
    pages = "25--32",
    year = "2002"
}

@article{Hambye:2003rt,
    author = "Hambye, Thomas and Lin, Yin and Notari, Alessio and Papucci, Michele and Strumia, Alessandro",
    title = "{Constraints on neutrino masses from leptogenesis models}",
    eprint = "hep-ph/0312203",
    archivePrefix = "arXiv",
    reportNumber = "IFUP-TH-2003-48",
    doi = "10.1016/j.nuclphysb.2004.06.027",
    journal = "Nucl. Phys. B",
    volume = "695",
    pages = "169--191",
    year = "2004"
}

@article{Abhishek:2026hex,
    author = "Abhishek and Mummidi, V. Suryanarayana",
    title = "{Asymmetric dark matter from leptogenesis in type-III seesaw framework with modular $S_4$ symmetry}",
    eprint = "2602.03384",
    archivePrefix = "arXiv",
    primaryClass = "hep-ph",
    month = "2",
    year = "2026"
}

@article{Baldes:2025uwz,
    author = "Baldes, Iason",
    title = "{Leptogenesis with perturbations in type-II and type-III seesaw models}",
    eprint = "2503.02936",
    archivePrefix = "arXiv",
    primaryClass = "hep-ph",
    doi = "10.1140/epjc/s10052-025-14277-9",
    journal = "Eur. Phys. J. C",
    volume = "85",
    number = "5",
    pages = "535",
    year = "2025"
}

@article{Suematsu:2019kst,
    author = "Suematsu, Daijiro",
    title = "{Low scale leptogenesis in a hybrid model of the scotogenic type I and III seesaw models}",
    eprint = "1906.12008",
    archivePrefix = "arXiv",
    primaryClass = "hep-ph",
    reportNumber = "KANAZAWA-19-04",
    doi = "10.1103/PhysRevD.100.055008",
    journal = "Phys. Rev. D",
    volume = "100",
    number = "5",
    pages = "055008",
    year = "2019"
}

@article{Roshan:2026yon,
    author = "Roshan, Rishav",
    title = "{Imprint of domain wall annihilation on induced gravitational waves}",
    eprint = "2604.25726",
    archivePrefix = "arXiv",
    primaryClass = "hep-ph",
    month = "4",
    year = "2026"
}

@article{Gouttenoire:2025ofv,
    author = "Gouttenoire, Yann and King, Stephen F. and Roshan, Rishav and Wang, Xin and White, Graham and Yamazaki, Masahito",
    title = "{Cosmological consequences of domain walls biased by quantum gravity}",
    eprint = "2501.16414",
    archivePrefix = "arXiv",
    primaryClass = "hep-ph",
    doi = "10.1103/7zmx-v16z",
    journal = "Phys. Rev. D",
    volume = "112",
    number = "7",
    pages = "075007",
    year = "2025"
}

@article{Roshan:2024qnv,
    author = "Roshan, Rishav and White, Graham",
    title = "{Using gravitational waves to see the first second of the Universe}",
    eprint = "2401.04388",
    archivePrefix = "arXiv",
    primaryClass = "hep-ph",
    doi = "10.1103/RevModPhys.97.015001",
    journal = "Rev. Mod. Phys.",
    volume = "97",
    number = "1",
    pages = "015001",
    year = "2025"
}

@article{Bhattacharya:2023kws,
    author = "Bhattacharya, Subhaditya and Mondal, Niloy and Roshan, Rishav and Vatsyayan, Drona",
    title = "{Leptogenesis, dark matter and gravitational waves from discrete symmetry breaking}",
    eprint = "2312.15053",
    archivePrefix = "arXiv",
    primaryClass = "hep-ph",
    doi = "10.1088/1475-7516/2024/06/029",
    journal = "JCAP",
    volume = "06",
    pages = "029",
    year = "2024"
}

@article{Datta:2021elq,
    author = "Datta, Arghyajit and Roshan, Rishav and Sil, Arunansu",
    title = "{Imprint of the Seesaw Mechanism on Feebly Interacting Dark Matter and the Baryon Asymmetry}",
    eprint = "2104.02030",
    archivePrefix = "arXiv",
    primaryClass = "hep-ph",
    doi = "10.1103/PhysRevLett.127.231801",
    journal = "Phys. Rev. Lett.",
    volume = "127",
    number = "23",
    pages = "231801",
    year = "2021"
}

@article{Mahapatra:2023dbr,
    author = "Mahapatra, Satyabrata and Paul, Partha Kumar and Sahu, Narendra and Shukla, Prashant",
    title = "{Asymmetric long-lived dark matter and leptogenesis from the type-III seesaw framework}",
    eprint = "2305.11138",
    archivePrefix = "arXiv",
    primaryClass = "hep-ph",
    doi = "10.1103/PhysRevD.111.015043",
    journal = "Phys. Rev. D",
    volume = "111",
    number = "1",
    pages = "015043",
    year = "2025"
}

@article{Mishra:2020gxg,
    author = "Mishra, Subhasmita",
    title = "{Neutrino mixing and Leptogenesis with modular $S_3$ symmetry in the framework of type III seesaw}",
    eprint = "2008.02095",
    archivePrefix = "arXiv",
    primaryClass = "hep-ph",
    month = "8",
    year = "2020"
}

@article{Chaudhuri:2026dvc,
    author = "Chaudhuri, Avinanda",
    title = "{Resonant Leptogenesis in a Two-Triplet Type-II Seesaw: A Dynamical Origin of Suppressed Lepton Flavor Violation}",
    eprint = "2604.11002",
    archivePrefix = "arXiv",
    primaryClass = "hep-ph",
    month = "4",
    year = "2026"
}

@article{Singh:2023eye,
    author = "Singh, Labh and Mahanta, Devabrat and Verma, Surender",
    title = "{Low scale leptogenesis in singlet-triplet scotogenic model}",
    eprint = "2309.12755",
    archivePrefix = "arXiv",
    primaryClass = "hep-ph",
    doi = "10.1088/1475-7516/2024/02/041",
    journal = "JCAP",
    volume = "02",
    pages = "041",
    year = "2024"
}

@article{Vatsyayan:2022rth,
    author = "Vatsyayan, Drona and Goswami, Srubabati",
    title = "{Lowering the scale of fermion triplet leptogenesis with two Higgs doublets}",
    eprint = "2208.12011",
    archivePrefix = "arXiv",
    primaryClass = "hep-ph",
    doi = "10.1103/PhysRevD.107.035014",
    journal = "Phys. Rev. D",
    volume = "107",
    number = "3",
    pages = "035014",
    year = "2023"
}

@article{Mishra:2022egy,
    author = "Mishra, Priya and Behera, Mitesh Kumar and Panda, Papia and Mohanta, Rukmani",
    title = "{Type III seesaw under $A_4$ modular symmetry with leptogenesis}",
    eprint = "2204.08338",
    archivePrefix = "arXiv",
    primaryClass = "hep-ph",
    doi = "10.1140/epjc/s10052-022-11074-6",
    journal = "Eur. Phys. J. C",
    volume = "82",
    number = "12",
    pages = "1115",
    year = "2022"
}

@article{Vilenkin:1984ib,
    author = "Vilenkin, Alexander",
    title = "{Cosmic Strings and Domain Walls}",
    reportNumber = "PRINT-84-0840 (TUFTS)",
    doi = "10.1016/0370-1573(85)90033-X",
    journal = "Phys. Rept.",
    volume = "121",
    pages = "263--315",
    year = "1985"
}

@article{Hiramatsu:2013qaa,
    author = "Hiramatsu, Takashi and Kawasaki, Masahiro and Saikawa, Ken'ichi",
    title = "{On the estimation of gravitational wave spectrum from cosmic domain walls}",
    eprint = "1309.5001",
    archivePrefix = "arXiv",
    primaryClass = "astro-ph.CO",
    reportNumber = "ICRR-REPORT-659-2013-8, IPMU13-0182, YITP-13-87",
    doi = "10.1088/1475-7516/2014/02/031",
    journal = "JCAP",
    volume = "02",
    pages = "031",
    year = "2014"
}

@article{Kitajima:2015nla,
    author = "Kitajima, Naoya and Takahashi, Fuminobu",
    title = "{Gravitational waves from Higgs domain walls}",
    eprint = "1502.03725",
    archivePrefix = "arXiv",
    primaryClass = "hep-ph",
    reportNumber = "TU-991, IPMU15-0016",
    doi = "10.1016/j.physletb.2015.04.040",
    journal = "Phys. Lett. B",
    volume = "745",
    pages = "112--117",
    year = "2015"
}

@article{Krajewski:2017czs,
    author = "Krajewski, Tomasz and Lalak, Zygmunt and Lewicki, Marek and Olszewski, Pawe{\l}",
    title = "{Domain walls in the extensions of the Standard Model}",
    eprint = "1709.10100",
    archivePrefix = "arXiv",
    primaryClass = "hep-ph",
    doi = "10.1088/1475-7516/2018/05/007",
    journal = "JCAP",
    volume = "05",
    pages = "007",
    year = "2018"
}

@article{Chen:2020wvu,
    author = "Chen, Ning and Li, Tong and Wu, Yongcheng",
    title = "{The gravitational waves from the collapsing domain walls in the complex singlet model}",
    eprint = "2004.10148",
    archivePrefix = "arXiv",
    primaryClass = "hep-ph",
    doi = "10.1007/JHEP08(2020)117",
    journal = "JHEP",
    volume = "08",
    pages = "117",
    year = "2020"
}

@article{Borah:2026kfo,
    author = "Borah, Debasish and Paul, Partha Kumar and Sahu, Narendra",
    title = "{Can Dirac neutrinos destabilize $\mathcal{Z}_2$ domain wall network?}",
    eprint = "2602.07380",
    archivePrefix = "arXiv",
    primaryClass = "hep-ph",
    month = "2",
    year = "2026"
}

@article{Paul:2024iie,
    author = "Paul, Partha Kumar and Sahu, Narendra and Shukla, Prashant",
    title = "{Thermal leptogenesis, dark matter, and gravitational waves from an extended canonical seesaw scenario}",
    eprint = "2409.08828",
    archivePrefix = "arXiv",
    primaryClass = "hep-ph",
    doi = "10.1103/w8gl-wbjd",
    journal = "Phys. Rev. D",
    volume = "112",
    number = "1",
    pages = "015032",
    year = "2025"
}

@article{Dey:2025pcs,
    author = "Dey, Ujjal Kumar and Manna, Santu Kumar and Paul, Partha Kumar and Sahoo, Sujit Kumar and Sahu, Narendra",
    title = "{Gravitational Wave Probe of Singlet-Doublet Dark Matter Induced Radiative Neutrino Mass}",
    eprint = "2511.19386",
    archivePrefix = "arXiv",
    primaryClass = "hep-ph",
    month = "11",
    year = "2025"
}

@article{Biswas:2025rzs,
    author = "Biswas, Anirban and Ganguly, Sougata",
    title = "{Probing low scale leptogenesis through gravitational wave}",
    eprint = "2505.01820",
    archivePrefix = "arXiv",
    primaryClass = "hep-ph",
    reportNumber = "CTPU-PTC-25-11",
    doi = "10.1103/zkrc-lrs5",
    journal = "Phys. Rev. D",
    volume = "113",
    number = "3",
    pages = "035028",
    year = "2026"
}

@article{Crowder:2005nr,
    author = "Crowder, Jeff and Cornish, Neil J.",
    title = "{Beyond LISA: Exploring future gravitational wave missions}",
    eprint = "gr-qc/0506015",
    archivePrefix = "arXiv",
    doi = "10.1103/PhysRevD.72.083005",
    journal = "Phys. Rev. D",
    volume = "72",
    pages = "083005",
    year = "2005"
}

@article{Reitze:2019iox,
    author = "Reitze, David and others",
    title = "{Cosmic Explorer: The U.S. Contribution to Gravitational-Wave Astronomy beyond LIGO}",
    eprint = "1907.04833",
    archivePrefix = "arXiv",
    primaryClass = "astro-ph.IM",
    reportNumber = "LIGO-P1900316",
    journal = "Bull. Am. Astron. Soc.",
    volume = "51",
    number = "7",
    pages = "035",
    year = "2019"
}

@article{Hild:2010id,
    author = "Hild, S. and others",
    title = "{Sensitivity Studies for Third-Generation Gravitational Wave Observatories}",
    eprint = "1012.0908",
    archivePrefix = "arXiv",
    primaryClass = "gr-qc",
    doi = "10.1088/0264-9381/28/9/094013",
    journal = "Class. Quant. Grav.",
    volume = "28",
    pages = "094013",
    year = "2011"
}

@article{Chen:2026fod,
    author = "Chen, Mu-Chun and Matias, Harold J. and Moffett-Smith, Cameron",
    title = "{Gravitational Waves in an A4 Neutrino Mass Model}",
    eprint = "2601.14394",
    archivePrefix = "arXiv",
    primaryClass = "hep-ph",
    month = "1",
    year = "2026"
}

@article{Sathyaprakash:2012jk,
    author = "Sathyaprakash, B. and others",
    editor = "Hannam, Mark and Sutton, Patrick and Hild, Stefan and van den Broeck, Chris",
    title = "{Scientific Objectives of Einstein Telescope}",
    eprint = "1206.0331",
    archivePrefix = "arXiv",
    primaryClass = "gr-qc",
    doi = "10.1088/0264-9381/29/12/124013",
    journal = "Class. Quant. Grav.",
    volume = "29",
    pages = "124013",
    year = "2012",
    note = "[Erratum: Class.Quant.Grav. 30, 079501 (2013)]"
}

@article{Punturo:2010zz,
    author = "Punturo, M. and others",
    editor = "Ricci, Fulvio",
    title = "{The Einstein Telescope: A third-generation gravitational wave observatory}",
    doi = "10.1088/0264-9381/27/19/194002",
    journal = "Class. Quant. Grav.",
    volume = "27",
    pages = "194002",
    year = "2010"
}

@article{LIGOScientific:2016aoc,
    author = "Abbott, B. P. and others",
    collaboration = "LIGO Scientific, Virgo",
    title = "{Observation of Gravitational Waves from a Binary Black Hole Merger}",
    eprint = "1602.03837",
    archivePrefix = "arXiv",
    primaryClass = "gr-qc",
    reportNumber = "LIGO-P150914",
    doi = "10.1103/PhysRevLett.116.061102",
    journal = "Phys. Rev. Lett.",
    volume = "116",
    number = "6",
    pages = "061102",
    year = "2016"
}

@article{LIGOScientific:2017vwq,
    author = "Abbott, B. P. and others",
    collaboration = "LIGO Scientific, Virgo",
    title = "{GW170817: Observation of Gravitational Waves from a Binary Neutron Star Inspiral}",
    eprint = "1710.05832",
    archivePrefix = "arXiv",
    primaryClass = "gr-qc",
    reportNumber = "LIGO-P170817",
    doi = "10.1103/PhysRevLett.119.161101",
    journal = "Phys. Rev. Lett.",
    volume = "119",
    number = "16",
    pages = "161101",
    year = "2017"
}

@article{LIGOScientific:2016wof,
    author = "Abbott, Benjamin P and others",
    collaboration = "LIGO Scientific",
    title = "{Exploring the Sensitivity of Next Generation Gravitational Wave Detectors}",
    eprint = "1607.08697",
    archivePrefix = "arXiv",
    primaryClass = "astro-ph.IM",
    reportNumber = "LIGO-P1600143",
    doi = "10.1088/1361-6382/aa51f4",
    journal = "Class. Quant. Grav.",
    volume = "34",
    number = "4",
    pages = "044001",
    year = "2017"
}

@article{Xing:2007fb,
    author = "Xing, Zhi-zhong and Zhang, He and Zhou, Shun",
    title = "{Updated Values of Running Quark and Lepton Masses}",
    eprint = "0712.1419",
    archivePrefix = "arXiv",
    primaryClass = "hep-ph",
    doi = "10.1103/PhysRevD.77.113016",
    journal = "Phys. Rev. D",
    volume = "77",
    pages = "113016",
    year = "2008"
}

@article{nEXO:2021ujk,
    author = "Adhikari, G. and others",
    collaboration = "nEXO",
    title = "{nEXO: neutrinoless double beta decay search beyond 10$^{28}$ year half-life sensitivity}",
    eprint = "2106.16243",
    archivePrefix = "arXiv",
    primaryClass = "nucl-ex",
    doi = "10.1088/1361-6471/ac3631",
    journal = "J. Phys. G",
    volume = "49",
    number = "1",
    pages = "015104",
    year = "2022"
}

@article{Planck:2018vyg,
    author = "Aghanim, N. and others",
    collaboration = "Planck",
    title = "{Planck 2018 results. VI. Cosmological parameters}",
    eprint = "1807.06209",
    archivePrefix = "arXiv",
    primaryClass = "astro-ph.CO",
    doi = "10.1051/0004-6361/201833910",
    journal = "Astron. Astrophys.",
    volume = "641",
    pages = "A6",
    year = "2020",
    note = "[Erratum: Astron.Astrophys. 652, C4 (2021)]"
}

@article{NuFit61,
    title = "{v6.1: Three-neutrino fit based on data available in November 2025}",
    doi = "http://www.nu-fit.org/?q=node/309",
    journal = "NuFit-6.1",
    year = "2025"
}

@article{ParticleDataGroup:2024cfk,
    author = "Navas, S. and others",
    collaboration = "Particle Data Group",
    title = "{Review of particle physics}",
    doi = "10.1103/PhysRevD.110.030001",
    journal = "Phys. Rev. D",
    volume = "110",
    number = "3",
    pages = "030001",
    year = "2024"
}

@article{Marciano:2024nwm,
    author = "Marciano, Simone and Meloni, Davide and Parriciatu, Matteo",
    title = "{Minimal seesaw and leptogenesis with the smallest modular finite group}",
    eprint = "2402.18547",
    archivePrefix = "arXiv",
    primaryClass = "hep-ph",
    doi = "10.1007/JHEP05(2024)020",
    journal = "JHEP",
    volume = "05",
    pages = "020",
    year = "2024"
}

@article{Priya:2025wdm,
    author = "Priya and Singh, Labh and Chauhan, B. C. and Verma, Surender",
    title = "{Type-III seesaw in non-holomorphic modular symmetry and leptogenesis}",
    eprint = "2508.05047",
    archivePrefix = "arXiv",
    primaryClass = "hep-ph",
    doi = "10.1007/JHEP01(2026)036",
    journal = "JHEP",
    volume = "01",
    pages = "036",
    year = "2026"
}

@article{Davidson:2008bu,
    author = "Davidson, Sacha and Nardi, Enrico and Nir, Yosef",
    title = "{Leptogenesis}",
    eprint = "0802.2962",
    archivePrefix = "arXiv",
    primaryClass = "hep-ph",
    doi = "10.1016/j.physrep.2008.06.002",
    journal = "Phys. Rept.",
    volume = "466",
    pages = "105--177",
    year = "2008"
}

@article{Tapender:2023kdk,
    author = "Tapender and Kumar, Sanjeev and Verma, Surender",
    title = "{Neutrino phenomenology in a model with generalized CP symmetry within type-I seesaw framework}",
    eprint = "2309.04242",
    archivePrefix = "arXiv",
    primaryClass = "hep-ph",
    doi = "10.1103/PhysRevD.109.015004",
    journal = "Phys. Rev. D",
    volume = "109",
    number = "1",
    pages = "015004",
    year = "2024"
}

@article{Barman:2025bru,
    author = "Barman, Animesh",
    title = "{Phenomenology of Neutrino Masses and Mixing with Discrete Flavour Symmetry in the context of the latest neutrino oscillation data}",
    school = "Tezpur University, Department of Physics, India",
    year = "2025"
}

@article{Chauhan:2023faf,
    author = "Chauhan, Garv and Dev, P. S. Bhupal and Dubovyk, Ievgen and Dziewit, Bartosz and Flieger, Wojciech and Grzanka, Krzysztof and Gluza, Janusz and Karmakar, Biswajit and Zi{\k{e}}ba, Szymon",
    title = "{Phenomenology of lepton masses and mixing with discrete flavor symmetries}",
    eprint = "2310.20681",
    archivePrefix = "arXiv",
    primaryClass = "hep-ph",
    doi = "10.1016/j.ppnp.2024.104126",
    journal = "Prog. Part. Nucl. Phys.",
    volume = "138",
    pages = "104126",
    year = "2024"
}

@article{Priya:2025khf,
    author = "Priya and Arora, Simran and Chauhan, B. C.",
    title = "{On the Generalized CP Symmetry, One Zero Texture in Neutrino Mass Matrix and Neutrinoless Double Beta Decay}",
    eprint = "2501.00776",
    archivePrefix = "arXiv",
    primaryClass = "hep-ph",
    month = "1",
    year = "2025"
}

@article{Ma:2025bjf,
    author = "Ma, Ernest and Paul, Partha Kumar and Sahu, Narendra",
    title = "{Lepton parity dark matter and naturally unstable domain walls}",
    eprint = "2508.02642",
    archivePrefix = "arXiv",
    primaryClass = "hep-ph",
    doi = "10.1103/tj6t-dyqn",
    journal = "Phys. Rev. D",
    volume = "112",
    number = "9",
    pages = "095020",
    year = "2025"
}

@article{Minkowski:1977sc,
    author = "Minkowski, Peter",
    title = "{$\mu \to e\gamma$ at a Rate of One Out of $10^{9}$ Muon Decays?}",
    reportNumber = "Print-77-0182 (BERN)",
    doi = "10.1016/0370-2693(77)90435-X",
    journal = "Phys. Lett. B",
    volume = "67",
    pages = "421--428",
    year = "1977"
}

@article{Mohapatra:1979ia,
    author = "Mohapatra, Rabindra N. and Senjanovic, Goran",
    title = "{Neutrino Mass and Spontaneous Parity Nonconservation}",
    reportNumber = "MDDP-TR-80-060, MDDP-PP-80-105, CCNY-HEP-79-10",
    doi = "10.1103/PhysRevLett.44.912",
    journal = "Phys. Rev. Lett.",
    volume = "44",
    pages = "912",
    year = "1980"
}

@article{Foot:1988aq,
    author = "Foot, Robert and Lew, H. and He, X. G. and Joshi, Girish C.",
    title = "{Seesaw Neutrino Masses Induced by a Triplet of Leptons}",
    reportNumber = "UM-P-88/89, OZ-P-88/7",
    doi = "10.1007/BF01415558",
    journal = "Z. Phys. C",
    volume = "44",
    pages = "441",
    year = "1989"
}

@article{Super-Kamiokande:1998kpq,
    author = "Fukuda, Y. and others",
    collaboration = "Super-Kamiokande",
    title = "{Evidence for oscillation of atmospheric neutrinos}",
    eprint = "hep-ex/9807003",
    archivePrefix = "arXiv",
    reportNumber = "BU-98-17, ICRR-REPORT-422-98-18, UCI-98-8, KEK-PREPRINT-98-95, LSU-HEPA-5-98, UMD-98-003, SBHEP-98-5, TKU-PAP-98-06, TIT-HPE-98-09",
    doi = "10.1103/PhysRevLett.81.1562",
    journal = "Phys. Rev. Lett.",
    volume = "81",
    pages = "1562--1567",
    year = "1998"
}

@article{SNO:2002tuh,
    author = "Ahmad, Q. R. and others",
    collaboration = "SNO",
    title = "{Direct evidence for neutrino flavor transformation from neutral current interactions in the Sudbury Neutrino Observatory}",
    eprint = "nucl-ex/0204008",
    archivePrefix = "arXiv",
    doi = "10.1103/PhysRevLett.89.011301",
    journal = "Phys. Rev. Lett.",
    volume = "89",
    pages = "011301",
    year = "2002"
}

@article{KamLAND:2002uet,
    author = "Eguchi, K. and others",
    collaboration = "KamLAND",
    title = "{First results from KamLAND: Evidence for reactor anti-neutrino disappearance}",
    eprint = "hep-ex/0212021",
    archivePrefix = "arXiv",
    doi = "10.1103/PhysRevLett.90.021802",
    journal = "Phys. Rev. Lett.",
    volume = "90",
    pages = "021802",
    year = "2003"
}

@article{DESI:2024mwx,
    author = "Adame, A. G. and others",
    collaboration = "DESI",
    title = "{DESI 2024 VI: cosmological constraints from the measurements of baryon acoustic oscillations}",
    eprint = "2404.03002",
    archivePrefix = "arXiv",
    primaryClass = "astro-ph.CO",
    reportNumber = "FERMILAB-PUB-24-0154-PPD",
    doi = "10.1088/1475-7516/2025/02/021",
    journal = "JCAP",
    volume = "02",
    pages = "021",
    year = "2025"
}

@article{LISA:2017pwj,
    author = "Amaro-Seoane, Pau and others",
    collaboration = "LISA",
    title = "{Laser Interferometer Space Antenna}",
    eprint = "1702.00786",
    archivePrefix = "arXiv",
    primaryClass = "astro-ph.IM",
    month = "2",
    year = "2017"
}

@article{TianQin:2015yph,
    author = "Luo, Jun and others",
    collaboration = "TianQin",
    title = "{TianQin: a space-borne gravitational wave detector}",
    eprint = "1512.02076",
    archivePrefix = "arXiv",
    primaryClass = "astro-ph.IM",
    doi = "10.1088/0264-9381/33/3/035010",
    journal = "Class. Quant. Grav.",
    volume = "33",
    number = "3",
    pages = "035010",
    year = "2016"
}

@article{TianQin:2020hid,
    author = "Mei, Jianwei and others",
    collaboration = "TianQin",
    title = "{The TianQin project: current progress on science and technology}",
    eprint = "2008.10332",
    archivePrefix = "arXiv",
    primaryClass = "gr-qc",
    doi = "10.1093/ptep/ptaa114",
    journal = "PTEP",
    volume = "2021",
    number = "5",
    pages = "05A107",
    year = "2021"
}

@article{Hu:2017mde,
    author = "Hu, Wen-Rui and Wu, Yue-Liang",
    title = "{The Taiji Program in Space for gravitational wave physics and the nature of gravity}",
    doi = "10.1093/nsr/nwx116",
    journal = "Natl. Sci. Rev.",
    volume = "4",
    number = "5",
    pages = "685--686",
    year = "2017"
}

@article{Ruan:2018tsw,
    author = "Ruan, Wen-Hong and Guo, Zong-Kuan and Cai, Rong-Gen and Zhang, Yuan-Zhong",
    title = "{Taiji program: Gravitational-wave sources}",
    eprint = "1807.09495",
    archivePrefix = "arXiv",
    primaryClass = "gr-qc",
    doi = "10.1142/S0217751X2050075X",
    journal = "Int. J. Mod. Phys. A",
    volume = "35",
    number = "17",
    pages = "2050075",
    year = "2020"
}

@article{LIGOScientific:2014qfs,
    author = "Aasi, J. and others",
    collaboration = "LIGO Scientific, VIRGO",
    title = "{Characterization of the LIGO detectors during their sixth science run}",
    eprint = "1410.7764",
    archivePrefix = "arXiv",
    primaryClass = "gr-qc",
    doi = "10.1088/0264-9381/32/11/115012",
    journal = "Class. Quant. Grav.",
    volume = "32",
    number = "11",
    pages = "115012",
    year = "2015"
}

@article{Esteban:2024eli,
    author = "Esteban, Ivan and Gonzalez-Garcia, M. C. and Maltoni, Michele and Martinez-Soler, Ivan and Pinheiro, Jo{\~a}o Paulo and Schwetz, Thomas",
    title = "{NuFit-6.0: updated global analysis of three-flavor neutrino oscillations}",
    eprint = "2410.05380",
    archivePrefix = "arXiv",
    primaryClass = "hep-ph",
    reportNumber = "IFT-UAM/CSIC-24-140, YITP-SB-2024-24, IPPP/24/64, IPPP/24/64, IFT-UAM/CSIC-24-140, YITP-SB-2024-24",
    doi = "10.1007/JHEP12(2024)216",
    journal = "JHEP",
    volume = "12",
    pages = "216",
    year = "2024"
}

@article{Hiramatsu:2010yz,
    author = "Hiramatsu, Takashi and Kawasaki, Masahiro and Saikawa, Ken'ichi",
    title = "{Gravitational Waves from Collapsing Domain Walls}",
    eprint = "1002.1555",
    archivePrefix = "arXiv",
    primaryClass = "astro-ph.CO",
    reportNumber = "ICRR-REPORT-559-2009-21, IPMU10-0024",
    doi = "10.1088/1475-7516/2010/05/032",
    journal = "JCAP",
    volume = "05",
    pages = "032",
    year = "2010"
}

@article{2852068,
    author = "Singh, Ishwar and Choudhary, Brajesh C. and Suter, Louise",
    collaboration = "NOvA",
    title = "{Latest Three-Flavor Neutrino Oscillation Results from NOvA}",
    reportNumber = "FERMILAB-CONF-24-0792-PPD"
}

@article{DUNE:2020fgq,
    author = "Abi, B. and others",
    collaboration = "DUNE",
    title = "{Prospects for beyond the Standard Model physics searches at the Deep Underground Neutrino Experiment}",
    eprint = "2008.12769",
    archivePrefix = "arXiv",
    primaryClass = "hep-ex",
    reportNumber = "FERMILAB-PUB-20-459-LBNF-ND, FERMILAB-PUB-20-459-LBNF-ND",
    doi = "10.1140/epjc/s10052-021-09007-w",
    journal = "Eur. Phys. J. C",
    volume = "81",
    number = "4",
    pages = "322",
    year = "2021"
}

@article{Hyper-Kamiokande:2018ofw,
    author = "Abe, K. and others",
    collaboration = "Hyper-Kamiokande",
    title = "{Hyper-Kamiokande Design Report}",
    eprint = "1805.04163",
    archivePrefix = "arXiv",
    primaryClass = "physics.ins-det",
    month = "5",
    year = "2018"
}

@article{Saikawa:2017hiv,
    author = "Saikawa, Ken'ichi",
    title = "{A review of gravitational waves from cosmic domain walls}",
    eprint = "1703.02576",
    archivePrefix = "arXiv",
    primaryClass = "hep-ph",
    reportNumber = "DESY-17-036",
    doi = "10.3390/universe3020040",
    journal = "Universe",
    volume = "3",
    number = "2",
    pages = "40",
    year = "2017"
}
\end{document}